\PassOptionsToPackage{unicode}{hyperref}
\PassOptionsToPackage{hyphens}{url}
\PassOptionsToPackage{dvipsnames,svgnames,x11names}{xcolor}
\documentclass[
  article]{article}
\usepackage{xcolor}
\usepackage{amsmath,amssymb}
  \usepackage[T1]{fontenc}
  \usepackage[utf8]{inputenc}
  \usepackage{textcomp} 
\usepackage{lmodern}
\IfFileExists{upquote.sty}{\usepackage{upquote}}{}
\IfFileExists{microtype.sty}{
  \usepackage[]{microtype}
  \UseMicrotypeSet[protrusion]{basicmath} 
}{}
\makeatletter
\@ifundefined{KOMAClassName}{
  \IfFileExists{parskip.sty}{%
    \usepackage{parskip}
  }{
    \setlength{\parindent}{0pt}
    \setlength{\parskip}{6pt plus 2pt minus 1pt}}
}{
  \KOMAoptions{parskip=half}}
\makeatother
\makeatletter
\ifx\paragraph\undefined\else
  \let\oldparagraph\paragraph
  \renewcommand{\paragraph}{
    \@ifstar
      \xxxParagraphStar
      \xxxParagraphNoStar
  }
  \newcommand{\xxxParagraphStar}[1]{\oldparagraph*{#1}\mbox{}}
  \newcommand{\xxxParagraphNoStar}[1]{\oldparagraph{#1}\mbox{}}
\fi
\ifx\subparagraph\undefined\else
  \let\oldsubparagraph\subparagraph
  \renewcommand{\subparagraph}{
    \@ifstar
      \xxxSubParagraphStar
      \xxxSubParagraphNoStar
  }
  \newcommand{\xxxSubParagraphStar}[1]{\oldsubparagraph*{#1}\mbox{}}
  \newcommand{\xxxSubParagraphNoStar}[1]{\oldsubparagraph{#1}\mbox{}}
\fi
\makeatother

\usepackage{color}
\usepackage{fancyvrb}

\DefineVerbatimEnvironment{Highlighting}{Verbatim}{commandchars=\\\{\}}
\newenvironment{Shaded}{}{}

\newcommand{\BuiltInTok}[1]{\textcolor[rgb]{0.65,0.15,0.64}{#1}}

\newcommand{\CommentTok}[1]{\textcolor[rgb]{0.63,0.63,0.65}{\textit{#1}}}

\newcommand{\ConstantTok}[1]{\textcolor[rgb]{0.60,0.41,0.00}{#1}}

\newcommand{\FloatTok}[1]{\textcolor[rgb]{0.60,0.41,0.00}{#1}}
\newcommand{\FunctionTok}[1]{\textcolor[rgb]{0.25,0.47,0.95}{#1}}
\newcommand{\ImportTok}[1]{\textcolor[rgb]{0.31,0.63,0.31}{#1}}

\newcommand{\KeywordTok}[1]{\textcolor[rgb]{0.65,0.15,0.64}{#1}}
\newcommand{\NormalTok}[1]{\textcolor[rgb]{0.22,0.23,0.26}{#1}}
\newcommand{\OperatorTok}[1]{\textcolor[rgb]{0.65,0.15,0.64}{#1}}

\newcommand{\PreprocessorTok}[1]{\textcolor[rgb]{0.65,0.15,0.64}{#1}}

\newcommand{\StringTok}[1]{\textcolor[rgb]{0.31,0.63,0.31}{#1}}

\usepackage{longtable,booktabs,array}
\usepackage{calc} 
\usepackage{etoolbox}
\makeatletter
\patchcmd\longtable{\par}{\if@noskipsec\mbox{}\fi\par}{}{}
\makeatother
\IfFileExists{footnotehyper.sty}{\usepackage{footnotehyper}}{\usepackage{footnote}}
\makesavenoteenv{longtable}
\usepackage{graphicx}
\makeatletter
\newsavebox\pandoc@box
\newcommand*\pandocbounded[1]{
  \sbox\pandoc@box{#1}%
  \Gscale@div\@tempa{\textheight}{\dimexpr\ht\pandoc@box+\dp\pandoc@box\relax}%
  \Gscale@div\@tempb{\linewidth}{\wd\pandoc@box}%
  \ifdim\@tempb\p@<\@tempa\p@\let\@tempa\@tempb\fi
  \ifdim\@tempa\p@<\p@\scalebox{\@tempa}{\usebox\pandoc@box}%
  \else\usebox{\pandoc@box}%
  \fi%
}
\def\fps@figure{htbp}
\makeatother

\usepackage[round]{natbib}
    \usepackage[T1]{fontenc}
    \usepackage[utf8]{inputenc}
    \usepackage{textalpha}
    \usepackage{bm}
    \usepackage[scale=0.8]{FiraMono}
    \usepackage{amsmath}
    \newcommand\mbfZ{{\mathbf{Z}}}
    \newcommand\mbfA{{\mathbf{A}}}
    \newcommand\mbfI{{\mathbf{I}}}
    \newcommand\mbfL{{\mathbf{L}}}
    
    \newcommand\mbfR{{\mathbf{R}}}
    
    \newcommand\mbfT{{\mathbf{T}}}
    \newcommand\mbfX{{\mathbf{X}}}
    \newcommand\mbfb{{\mathbf{b}}}
    
    \newcommand\mbfu{{\mathbf{u}}}
    \newcommand\mbfv{{\mathbf{v}}}
    \newcommand\mbfy{{\mathbf{y}}}
    \newcommand\mcN{{\bm{\mathcal{N}}}}
    \newcommand\mcB{{\bm{\mathcal{B}}}}
    \newcommand\mcU{{\bm{\mathcal{U}}}}
    \newcommand\mcY{{\bm{\mathcal{Y}}}}
    \newcommand\mbfLambda{{\boldsymbol{\Lambda}}}
    \newcommand\mbfOmega{{\boldsymbol{\Omega}}}
    \newcommand\mbfSigma{{\boldsymbol{\Sigma}}}
    \newcommand\mbfbeta{{\boldsymbol{\beta}}}
    \newcommand\mbftheta{{\boldsymbol{\theta}}}
    \DeclareUnicodeCharacter{2500}{\textemdash} 
    \DeclareUnicodeCharacter{0302}{\kern-0.6em\raisebox{0.5ex}{\textasciicircum}} 
    \DeclareUnicodeCharacter{22C5}{.}
    \DeclareUnicodeCharacter{22F1}{*}
    \DeclareUnicodeCharacter{2265}{$\geq$}
    \DeclareUnicodeCharacter{251C}{$\vdash$} 
    \DeclareUnicodeCharacter{2514}{\linebreak{}$\llcorner$} 
    \DeclareUnicodeCharacter{2303}{\^}

\usepackage{graphicx}
\usepackage{arxiv}
\usepackage{orcidlink}
\usepackage{amsmath}
\usepackage{lmodern}
\usepackage[T1]{fontenc}
\let\proglang=\texttt
\newcommand{\pkg}[1]{{\fontseries{m}\fontseries{b}\selectfont #1}}
\makeatletter
\@ifpackageloaded{caption}{}{\usepackage{caption}}
\AtBeginDocument{%
\ifdefined\contentsname
  \renewcommand*\contentsname{Table of contents}
\else
  \newcommand\contentsname{Table of contents}
\fi
\ifdefined\listfigurename
  \renewcommand*\listfigurename{List of Figures}
\else
  \newcommand\listfigurename{List of Figures}
\fi
\ifdefined\listtablename
  \renewcommand*\listtablename{List of Tables}
\else
  \newcommand\listtablename{List of Tables}
\fi
\ifdefined\figurename
  \renewcommand*\figurename{Figure}
\else
  \newcommand\figurename{Figure}
\fi
\ifdefined\tablename
  \renewcommand*\tablename{Table}
\else
  \newcommand\tablename{Table}
\fi
}
\@ifpackageloaded{float}{}{\usepackage{float}}
\floatstyle{ruled}
\@ifundefined{c@chapter}{\newfloat{codelisting}{h}{lop}}{\newfloat{codelisting}{h}{lop}[chapter]}
\floatname{codelisting}{Listing}

\makeatother
\makeatletter
\@ifpackageloaded{caption}{}{\usepackage{caption}}
\@ifpackageloaded{subcaption}{}{\usepackage{subcaption}}
\makeatother
\usepackage{bookmark}
\IfFileExists{xurl.sty}{\usepackage{xurl}}{} 
\hypersetup{
  pdftitle={Mixed-model Log-likelihood Evaluation Via a Blocked Cholesky Factorization},
  pdfauthor={Douglas Bates; Phillip M. Alday; Ajinkya H. Kokandakar},
  colorlinks=true,
  linkcolor={blue},
  filecolor={Maroon},
  citecolor={Blue},
  urlcolor={Blue},
  pdfcreator={LaTeX via pandoc}}

\newcommand{\runninghead}{A Preprint }
\renewcommand{\runninghead}{Mixed Models in Julia }
\title{Mixed-model Log-likelihood Evaluation Via a Blocked Cholesky
Factorization}
\def\asep{\\\\\\ } 
\author{\textbf{Douglas
Bates}~\orcidlink{0000-0001-8316-9503}\\Department of
Statistics\\University of
Wisconsin--Madison\\\\\href{mailto:dmbates@gmail.com}{dmbates@gmail.com}\asep\textbf{Phillip
M. Alday}~\orcidlink{0000-0002-9984-5745}\\\\Beacon
Biosignals\\\\\href{mailto:phillip@phillipalday.com}{phillip@phillipalday.com}\asep\textbf{Ajinkya
H. Kokandakar}~\orcidlink{0000-0001-6628-2272}\\Department of
Statistics\\University of
Wisconsin--Madison\\\\\href{mailto:ajinkya@stat.wisc.edu}{ajinkya@stat.wisc.edu}}
\date{15 May 2025}

\begin{document}
\maketitle
\begin{abstract}
\citet{bates.maechler.etal:2015} described the evaluation of the
profiled log-likelihood of a linear mixed-effects model by updating a
sparse, symmetric positive-definite matrix and computing its Cholesky
factor, as implemented in the \pkg{lme4} package for \proglang{R}. Here
we present enhancements to the derivation and theoretical presentation
of the result and to its implementation using a blocked Cholesky
factorization in the \pkg{MixedModels.jl} package for \proglang{Julia}
\citep{Bezanson2017}. The gain in computational efficiency is primarily
due to three factors: (1) the new derivation allows us to compute the
penalized residual sum of squares without computing the conditional
estimates of the fixed-effects parameters and the conditional modes of
the random effects at each optimization step, (2) the blocked Cholesky
representation and careful ordering of the random effects terms reduces
the amount of ``fill-in'' that occurs during the Cholesky factorization,
and (3) the multiple dispatch feature of the \proglang{Julia} language
allows us to use specialized algorithms for different kinds of matrices
instead of relying on generic algorithms during the Cholesky
factorization. To show the effectiveness of the blocked Cholesky
approach we use it to fit a linear mixed model to over 32 million
ratings of movies in the MovieLens \texttt{ml-32m} \citep{Harper2016}
data set. The model incorporates random effects for over 200,000 movies
and over 80,000 participants. Further enhancements to these
computational methods are suggested.
\end{abstract}

\section{Introduction}\label{sec-intro}

Mixed effects models are a rich class of statistical models frequently
used when the measurement units exhibit grouping structures. They are
especially useful when the data exhibit non-nested grouping, for
instance when there are two or more unrelated grouping variables such as
measurements taken on items and users
\citep[see][]{baayendavidsonbates2008a}; or partially nested groupings,
for instance when measuring test scores for students over time when the
students can change schools \citep[see][]{fuhner2021age}. There is a
long history of computational tools for fitting mixed effects models,
including the \texttt{PROC\ MIXED} procedure in \proglang{SAS}, and the
\pkg{nlme}, \pkg{lme4} and more recently \pkg{glmmTMB} packages in the
\proglang{R} ecosystem. The extensive use of these models is shown by
the fact that \citet{bates.maechler.etal:2015}, which describes the
methodology and use of \pkg{lme4}, is among the top 25 most-cited papers
of the twenty-first century \citep{Pearson2025}. We present
\pkg{MixedModels.jl} a new implementation of mixed effects models in the
\proglang{Julia} programming language. This implementation relies on two
innovations, one analytical and the other computational. At the heart of
\pkg{MixedModels.jl}, there is a new way of formulating the profiled
log-likelihood evaluation that expresses this objective without
requiring explicit evaluation of the conditional parameter estimates. On
the computational level, the evaluation of the objective requires the
Cholesky factor of a large, sparse, positive-definite symmetric matrix,
as in \pkg{lme4}. This matrix has a blocked structure that is explicitly
exploited in \pkg{MixedModels.jl} but not in \pkg{lme4}. More
concretely, the Cholesky factor consists of dense and sparse blocks, and
using specific algorithms to handle these blocks differently leads to
more computationally efficient algorithms. We exploit the multiple
dispatch feature of \proglang{Julia} to do exactly that -- use sparse
matrix algorithms for the sparse blocks, and level 3 BLAS and LAPACK
subroutines for the dense blocks. Together, these contributions allow us
to fit linear mixed models to very large datasets.

\citet{bates.maechler.etal:2015} provided a formulation for linear
mixed-effects models and showed how maximum likelihood estimates of the
parameters in the model can be determined by optimizing a profiled
objective function, which is negative twice the profiled log-likelihood.
The term ``profiled'' refers to the fact that this objective is a
function of some (but not all) of the parameters in the model. As shown
in \citet{bates.maechler.etal:2015}, the profiled objective can be
evaluated from the solution of a penalized least squares (PLS) problem.
In this paper, we show that it is not necessary to evaluate the solution
to the PLS problem explicitly when evaluating the profiled objective.
The objective can be evaluated from a factorization that is an
intermediate step in solving the PLS problem.

The model matrix, \(\mbfZ\), for the random effects in such models is
defined in blocks according to the ``grouping factors'' for
random-effects terms in the model formula. The blocked structure of
\(\mbfZ\) and of its cross-product matrix, \(\mbfZ^\top\mbfZ\), (also
called the Gram matrix of the columns of \(\mbfZ\)) is described in
great detail in \citet[Section 2]{bates.maechler.etal:2015} when showing
how a model formula and data frame create the structures used to
evaluate the profiled objective. However, the implementation of the
profiled log-likelihood evaluation in the \pkg{lme4} package for
\proglang{R} is not explicitly based upon the blocks --- \pkg{lme4} uses
a more general sparse Cholesky factorization as implemented in the
\pkg{CHOLMOD} library \citep{10.1145/1391989.1391995}.

An important part in a general sparse Cholesky factorization, such as
implemented in \pkg{CHOLMOD}, is determining a \emph{fill-reducing
permutation} of the rows and columns of the symmetric, positive-definite
matrix to be factored. The main difference between the methods in
\pkg{MixedModels.jl} and those in \pkg{lme4} is that in
\pkg{MixedModels.jl} it is the blocks that are permuted, if necessary,
rather than individual rows and columns. In other words, the particular
structure of the blocks in the Gram matrix is exploited in
\pkg{MixedModels.jl} rather than relying upon general sparse Cholesky
techniques that are not able to exploit this structure.

The structure of the blocks in the Gram matrix depends upon the number
and types of random-effects terms in the model. A general and efficient
implementation of the blocked Cholesky factorization requires the
ability to define operations for many different combinations of
specialized forms of matrices. \proglang{Julia} \citep{Bezanson2017} is
an ideal environment for this type of calculation because of its
efficient implementation of multiple dispatch, its just-in-time
compilation, and its rich type system.

The rest of the paper is organized as follows. In
Section~\ref{sec-profiled} we recap the derivation of the profiled
log-likelihood from \citet{bates.maechler.etal:2015}, in a slightly
different form from that paper, and show that the objective can be
evaluated directly from the Cholesky factorization of an augmented
positive-definite matrix. That is, explicit evaluation of the solution
to the PLS problem is not necessary. In Section~\ref{sec-structure} we
use an example of a fitted mixed-effects model to illustrate our
implementation of the blocked Cholesky factorization in
\pkg{MixedModels.jl}. In Section~\ref{sec-application} we use an example
of a mixed-effects model fit to a large data set with partially crossed
grouping factors for the random effects to examine how the compute time
per evaluation of the objective and the memory footprint of the model
depend on various dimensions of the model. We conclude with a few
additional comments in Section~\ref{sec-further-comments} followed by a
short summary and discussion in Section~\ref{sec-summary}.

\section{Profiled log-likelihood of linear mixed-effects
models}\label{sec-profiled}

The types of linear mixed-effects models that can be fit by \pkg{lme4}
or \pkg{MixedModels.jl} are based on two vector-valued random variables:
the \(q\)-dimensional vector of random effects, \(\mcB\), and the
\(n\)-dimensional response vector, \(\mcY\). The unconditional
distribution of \(\mcB\) and the conditional distribution of \(\mcY\),
given \(\mcB=\mathbf{b}\), are defined as multivariate Gaussian
distributions of the form
\begin{equation}\phantomsection\label{eq-LMMdistr}{
\begin{aligned}
(\mcY|\mcB=\mathbf{b})&\sim\mcN(\mbfX\mbfbeta+\mathbf{Z}\mathbf{b},\sigma^2\mbfI)\\
\mcB&\sim\mcN(\mathbf{0},\mbfSigma_\theta)\ ,
\end{aligned}
}\end{equation}

where \(\mbfX\) is the \(n\times p\) model matrix for the fixed-effects
terms, \(\mbfbeta\) is the \(p\)-dimensional fixed-effects parameter
vector and \(\mathbf{Z}\) is the \(n\times q\) model matrix for the
random effects. The \(q\times q\) symmetric variance-covariance matrix
\(\mathrm{Var}(\mcB)=\mbfSigma_\theta\) depends on the
\emph{variance-component parameter vector} \(\mbftheta\) through a lower
triangular \emph{relative covariance factor} \(\mbfLambda_\theta\) as

\begin{equation}\phantomsection\label{eq-relcovfac}{
\mbfSigma_\theta=\sigma^2\mbfLambda_\theta\mbfLambda_\theta^\top\ .
}\end{equation}

(Recall that the lower Cholesky factor is generally written
\(\mathbf{L}\). In this case the lower Cholesky factor contains
parameters and is named with the corresponding Greek letter,
\(\mbfLambda\).)

The matrices \(\mbfZ\), \(\mbfSigma_\theta\) and \(\mbfLambda_\theta\)
can be very large but have special sparsity structures, as described in
Section~\ref{sec-structure}.

In many descriptions of linear mixed models, computational formulas are
written in terms of the \emph{precision matrix},
\(\mbfSigma_\theta^{-1}\). Such formulas will become unstable as
\(\mbfSigma_\theta\) approaches singularity. And it can do so; singular
(i.e.~non-invertible) \(\mbfSigma_\theta\) can and do occur in practice.
Moreover, during the course of the numerical optimization by which the
parameter estimates are determined, it is frequently the case that the
log-likelihood or the REML criterion will need to be evaluated at values
of \(\mbftheta\) that produce a singular \(\mbfSigma_\theta\). Because
of this we will take care to use computational methods that can be
applied even when \(\mbfSigma_\theta\) is singular and are stable as
\(\mbfSigma_\theta\) approaches singularity.

According to Equation~\ref{eq-relcovfac}, \(\mbfSigma\) depends on both
\(\sigma\) and \(\theta\), and we should write it as
\(\mbfSigma_{\sigma,\theta}\). However, we will blur that distinction
and continue to write \(\text{Var}(\mcB)=\mbfSigma_\theta\).

Another technicality is that the \emph{common scale parameter},
\(\sigma\), could, in theory, be zero. However, the only way for its
estimate, \(\widehat{\sigma}\), to be zero is for the fitted values from
the fixed-effects only, \(\mbfX\widehat{\mbfbeta}\), to be exactly equal
to the observed data. This occurs only with data that have been
(incorrectly) simulated without error. In practice, we can safely assume
that \(\sigma>0\). However, the estimated \(\mbfLambda_\theta\), like
\(\mbfSigma_\theta\), can be singular.

The computational methods in \pkg{lme4} and in \pkg{MixedModels.jl} are
based on \(\mbfLambda_\theta\) and do not require evaluation of
\(\mbfSigma_\theta\). In fact, \(\mbfSigma_\theta\) is explicitly
evaluated only at the converged parameter estimates.

To express the log likelihood without the precision matrix, we define
the spherical random effects,
\(\mcU\sim\mcN(\mathbf{0},\sigma^2\mbfI_q)\), which in turn determine
\(\mcB\) as

\begin{equation}\phantomsection\label{eq-sphericalre}{
\mcB=\mbfLambda_\theta\mcU\ .
}\end{equation}

Although it may seem more intuitive to write \(\mcU\) as a linear
transformation of \(\mcB\), we cannot do that when \(\mbfLambda_\theta\)
is singular, which is why Equation~\ref{eq-sphericalre} is in the form
shown.

We can easily verify that Equation~\ref{eq-sphericalre} provides the
desired distribution for \(\mcB\). As a linear transformation of a
multivariate Gaussian random variable, \(\mcB\) will also be
multivariate Gaussian with mean

\[
\mathbb{E}\left[\mcB\right]=
\mathbb{E}\left[\mbfLambda_\theta\mcU\right]=
\mbfLambda_\theta\,\mathbb{E}\left[\mcU\right]=
\mbfLambda_\theta\mathbf{0}=\mathbf{0}\ ,
\]

and covariance matrix

\[
\text{Var}(\mcB)=
\mbfLambda_\theta\text{Var}(\mcU)\mbfLambda_\theta^{\top}=
\sigma^2\mbfLambda_\theta\mbfLambda_\theta^{\top}=\mbfSigma_\theta \ .
\]

Just as we concentrate on how \(\mbftheta\) determines
\(\mbfLambda_\theta\), not \(\mbfSigma_\theta\), we will concentrate on
properties of \(\mcU\) rather than \(\mcB\). In particular, we now
define the model according to the distributions

\begin{equation}\phantomsection\label{eq-condygivenu}{
\begin{aligned}
(\mcY|\mcU=\mathbf{u})&\sim\mcN(\mathbf{Z}\mbfLambda_\theta\mathbf{u}+\mbfX\mbfbeta,\sigma^2\mbfI_n)\\
\mcU&\sim\mcN(\mathbf{0},\sigma^2\mbfI_q)\ .
\end{aligned}
}\end{equation}

The joint density function for \(\mcY\) and \(\mcU\) is the product of
the densities of the two distributions in Equation~\ref{eq-condygivenu}.
That is

\begin{equation}\phantomsection\label{eq-yujointdensity}{
f_{\mcY,\mcU}(\mbfy,\mathbf{u})=
\frac{1}{\left(2\pi\sigma^2\right)^{(n+q)/2}}\exp
\left(\frac{\left\|\mbfy-\mbfX\mbfbeta
-\mathbf{Z}\mbfLambda_\theta\mathbf{u}\right\|^2+
\left\|\mathbf{u}\right\|^2}{-2\sigma^2}\right)\ .
}\end{equation}

To evaluate the likelihood for the parameters, \(\mbftheta\),
\(\mbfbeta\), and \(\sigma^2\), given the observed response, \(\mbfy\),
we must evaluate the marginal distribution of \(\mcY\), whose density is
the integral of \(f_{\mcY,\mcU}(\mbfy,\mathbf{u})\) with respect to
\(\mathbf{u}\).

This evaluation can be simplified if we rewrite the \emph{penalized sum
of squared residuals}, \(\left\|\mbfy-\mbfX\mbfbeta
-\mathbf{Z}\mbfLambda_\theta\mathbf{u}\right\|^2+
\left\|\mathbf{u}\right\|^2\), the numerator of the exponent in
Equation~\ref{eq-yujointdensity}, as a quadratic form in \(\mathbf{u}\),
to isolate the dependence on \(\mathbf{u}\)

\begin{equation}\phantomsection\label{eq-penalized-rss}{
\begin{aligned}
r^2_\theta(\mathbf{u},\mbfbeta)
&=
\|\mbfy-\mbfX\mbfbeta-\mathbf{Z}\mbfLambda_\theta\mathbf{u}\|^2+\|\mathbf{u}\|^2 \\
&=
\left\|
\begin{bmatrix}
\mathbf{Z}\mbfLambda_\theta & \mbfX & \mbfy \\
-\mbfI_q & \mathbf{0} & \mathbf{0}
\end{bmatrix}
\begin{bmatrix}
-\mathbf{u} \\
-\mbfbeta \\
1
\end{bmatrix}
\right\|^2 \\
&=
\begin{bmatrix}
-\mathbf{u} \\
-\mbfbeta \\
1
\end{bmatrix}^{\top}
\begin{bmatrix}
\mbfLambda_\theta^{\top}\mathbf{Z}^{\top}\mathbf{Z}\mbfLambda_\theta+\mbfI & \mbfLambda_\theta^{\top}\mathbf{Z}^{\top}\mbfX & \mbfLambda_\theta^{\top}\mathbf{Z}^{\top}\mbfy \\
\mbfX^{\top}\mathbf{Z}\mbfLambda_\theta & \mbfX^{\top}\mbfX & \mbfX^{\top}\mbfy \\
\mbfy^{\top}\mathbf{Z}\mbfLambda_\theta & \mbfy^{\top}\mbfX & \mbfy^{\top}\mbfy
\end{bmatrix}
\begin{bmatrix}
-\mathbf{u} \\
-\mbfbeta \\
1
\end{bmatrix} \\
&= \left\|
\begin{bmatrix}
\mbfR_{ZZ} & \mbfR_{ZX} & \mathbf{r}_{Zy}\\
\mathbf{0} & \mbfR_{XX} & \mathbf{r}_{Xy}\\
\mathbf{0} & \mathbf{0} & r_{yy}
\end{bmatrix}
\begin{bmatrix}
-\mathbf{u} \\
-\mbfbeta \\
1
\end{bmatrix}
\right\|^2\\
&= \| \mathbf{r}_{Zy}-\mbfR_{ZX}\mbfbeta-\mbfR_{ZZ}\mathbf{u} \|^2 +
\| \mathbf{r}_{Xy}-\mbfR_{XX}\mbfbeta\|^2 + r_{yy}^2\ ,
\end{aligned}
}\end{equation}

using the Cholesky factor of the blocked matrix,

\begin{equation}\phantomsection\label{eq-bigCholfac}{
\begin{aligned}
\mbfOmega_\theta&=
\begin{bmatrix}
\mbfLambda_\theta^{\top}\mathbf{Z^{\top}Z}\mbfLambda_\theta+\mbfI &
\mbfLambda_\theta^{\top}\mathbf{Z^{\top}X} & \mbfLambda_\theta^{\top}\mathbf{Z^{\top}y} \\
\mathbf{X^{\top}Z}\mbfLambda_\theta & \mathbf{X^{\top}X} & \mathbf{X^{\top}y} \\
\mathbf{y^{\top}Z}\mbfLambda_\theta & \mathbf{y^{\top}X} & \mathbf{y^{\top}y}
\end{bmatrix}\\
& =
\begin{bmatrix}
\mbfR_{ZZ}^{\top} & \mathbf{0} & \mathbf{0} \\
\mbfR_{ZX}^{\top} & \mbfR^{\top}_{XX} & \mathbf{0} \\
\mathbf{r}_{Zy}^{\top} & \mathbf{r}^{\top}_{Xy} & r_{yy}
\end{bmatrix}
\begin{bmatrix}
\mbfR_{ZZ} & \mbfR_{ZX} & \mathbf{r}_{Zy} \\
\mathbf{0} & \mbfR_{XX} & \mathbf{r}_{Xy} \\
\mathbf{0} & \mathbf{0} & r_{yy}
\end{bmatrix}\ .
\end{aligned}
}\end{equation}

(In practice we evaluate the blocks of the lower Cholesky factor, such
as \(\mbfL_{ZZ}=\mbfR_{ZZ}^{\top}\), of this matrix but the equations
are slightly easier to write in terms of the upper Cholesky factor.)

Note that the block in the upper left,
\(\mbfLambda_\theta^{\top}\mathbf{Z^{\top}Z}\mbfLambda_\theta+\mbfI\),
is positive definite even when \(\mbfLambda_\theta\) is singular,
because

\begin{equation}\phantomsection\label{eq-Cholfacupperleft}{
\mathbf{u}^{\top}\left(\mbfLambda_\theta^{\top}\mathbf{Z^{\top}Z}\mbfLambda_\theta+\mbfI\right)\mathbf{u} = \left\|\mathbf{Z}\mbfLambda_\theta\mathbf{u}\right\|^2
+\left\|\mathbf{u}\right\|^2\ ,
}\end{equation} and the first term is non-negative while the second is
positive if \(\mathbf{u}\ne\mathbf{0}\).

Thus, the triangular \(\mbfR_{ZZ}\), with positive diagonal elements,
can be evaluated and its determinant, \(\left|\mbfR_{ZZ}\right|\), which
is the product of its diagonal elements, is also positive. This
determinant appears in the marginal density of \(\mcY\), from which the
likelihood of the parameters is evaluated.

To evaluate the likelihood,

\begin{equation}\phantomsection\label{eq-likelihood-abstract}{
L(\mbftheta,\mbfbeta,\sigma|\mbfy) = \int_\mathbf{u} f_{\mcY,\mcU}(\mbfy,\mathbf{u})\, d\mathbf{u}\ ,
}\end{equation}

we isolate the part of the joint density that depends on \(\mathbf{u}\)
and perform a change of variable

\begin{equation}\phantomsection\label{eq-u-system}{
\mathbf{v}=\mbfR_{ZZ}\mathbf{u}+\mbfR_{ZX}\mbfbeta-\mathbf{r}_{Zy} \ ,
}\end{equation}

with the \emph{Jacobian matrix}

\[
\frac{d\mbfv}{d\mbfu} = \mbfR_{ZZ}\ .
\]

From the properties of the multivariate Gaussian distribution

\begin{equation}\phantomsection\label{eq-likelihood-integral}{
\begin{aligned}
\int_{\mathbf{u}}\frac{1}{(2\pi\sigma^2)^{q/2}}
\exp\left(-\frac{\|\mbfR_{ZZ}\mathbf{u}+\mbfR_{ZX}\mbfbeta-\mathbf{r}_{Zy}\|^2}{2\sigma^2}\right)
\,d\mathbf{u}\\
\begin{aligned}
&= \int_{\mathbf{v}}\frac{1}{(2\pi\sigma^2)^{q/2}}
\exp\left(-\frac{\|\mathbf{v}\|^2}{2\sigma^2}\right)|\mbfR_{ZZ}|^{-1}\,d\mathbf{v}\\
&=|\mbfR_{ZZ}|^{-1} \ ,
\end{aligned}
\end{aligned}
}\end{equation}

from which we obtain the likelihood as

\begin{equation}\phantomsection\label{eq-likelihood}{
L(\mbftheta,\mbfbeta,\sigma;\mbfy)=
\frac{|\mbfR_{ZZ}|^{-1}}{(2\pi\sigma^2)^{n/2}}
\exp\left(-\frac{r_{yy}^2 + \|\mbfR_{XX}(\mbfbeta-\widehat{\mbfbeta})\|^2}{2\sigma^2}\right)\ ,
}\end{equation}

where the conditional estimate, \(\widehat{\mbfbeta}\), given
\(\mbftheta\), satisfies \[
\mbfR_{XX}\widehat{\mbfbeta} = \mathbf{r}_{Xy}\ .
\]

Setting \(\mbfbeta=\widehat{\mbfbeta}\) and taking the logarithm
provides the estimate of \(\sigma^2\), given \(\mbftheta\), as

\begin{equation}\phantomsection\label{eq-sigma-hat}{
\widehat{\sigma^2}=\frac{r_yy^2}{n}\ ,
}\end{equation}

which gives the \emph{profiled log-likelihood},
\(\ell(\mbftheta|\mbfy)=\log L(\mbftheta,\widehat{\mbfbeta},\widehat{\sigma})\),
on the deviance scale, as

\begin{equation}\phantomsection\label{eq-profiled-log-likelihood}{
-2\ell(\mbftheta|\mbfy)=2\log(|\mbfR_{ZZ}|) +
n\left(1+\log\left(\frac{2\pi r_{yy}^2(\mbftheta)}{n}\right)\right)\ .
}\end{equation}

One of the interesting aspects of this formulation is that it is not
necessary to solve for the conditional estimate of \(\mbfbeta\) or the
conditional modes of the random effects when evaluating the
log-likelihood. The two values needed for the log-likelihood evaluation,
\(2\log(|\mbfR_{ZZ}|)\) and \(r_\mathbf{yy}^2\), are obtained directly
from the diagonal elements of the Cholesky factor.

Furthermore, \(\mbfOmega_{\theta}\) and, from that, the Cholesky factor,
\(\mbfR_{\theta}\), and the objective to be optimized can be evaluated
for a given value of \(\mbftheta\) from the Gram matrix of the columns
of \(\mbfZ\), \(\mbfX\) and \(\mbfy\),

\begin{equation}\phantomsection\label{eq-A}{
\mathbf{A} = \begin{bmatrix}
\mathbf{Z}^{\top}\mathbf{Z} & \mathbf{Z}^{\top}\mbfX & \mathbf{Z}^{\top}\mbfy \\
\mbfX^{\top}\mathbf{Z} & \mbfX^{\top}\mbfX & \mbfX^{\top}\mbfy \\
\mbfy^{\top}\mathbf{Z} & \mbfy^{\top}\mbfX & \mbfy^{\top}\mbfy
\end{bmatrix}\ ,
}\end{equation}

and \(\mbfLambda_{\theta}\).

In \pkg{MixedModels.jl} the \texttt{LinearMixedModel} struct contains a
blocked representation of \(\mathbf{A}\) in the \texttt{A} field and a
similarly structured lower-triangular blocked array in the \texttt{L}
field. Evaluation of the objective simply involves updating a
representation of the relative covariance factor \(\mbfLambda_\theta\)
from \(\mbftheta\), forming \(\mbfOmega_\theta\)
(Equation~\ref{eq-bigCholfac}) from \(\mbfA\) and \(\mbfLambda_\theta\)
then evaluating its lower Cholesky factor \(\mbfL_\theta\).

In Equation~\ref{eq-A} the blocking structure is according to all the
random effects, followed by the fixed-effects followed by the response.
However, as described in \citet[Section 3]{bates.maechler.etal:2015},
the structure of \(\mbfZ\), \(\mbfZ^{\top}\mbfZ\) and
\(\mbfLambda_\theta\) is further subdivided into blocks according to the
random-effects terms in the model formula, which we cover in the next
section.

\section{Random-effects terms in mixed-model
formulas}\label{sec-structure}

The user interface in \pkg{MixedModels.jl} is intentionally similar to
that of \pkg{lme4}. A model is specified, as described in \citet[section
2.1]{bates.maechler.etal:2015}, by a formula of the form

\begin{Shaded}
\begin{Highlighting}[]
\PreprocessorTok{@formula}\NormalTok{(resp }\OperatorTok{\textasciitilde{}}\NormalTok{ FEexpr }\OperatorTok{+}\NormalTok{ (REexpr1 }\OperatorTok{|}\NormalTok{ factor1) }\OperatorTok{+}\NormalTok{ (REexpr2 }\OperatorTok{|}\NormalTok{ factor2) }\OperatorTok{+} \OperatorTok{...}\NormalTok{)}
\end{Highlighting}
\end{Shaded}

``where \texttt{FEexpr} is an expression determining the columns of the
fixed-effects model matrix, \(\mbfX\), and the random-effects terms,
\texttt{(REexpr1\ \textbar{}\ factor1)},
\texttt{(REexpr2\ \textbar{}\ factor2)}, etc. together determine both
the random-effects model matrix \(\mbfZ\) and the structure of the
relative covariance factor \(\mbfLambda_\theta\).''

(In \proglang{Julia} a formula expression such as this must be enclosed
in a call to the \texttt{@formula} macro.)

The expression on the right hand side of the vertical bar,
\texttt{\textbar{}}, in a random-effects term is called the
\emph{grouping factor} for those random effects.

\subsection{From formula to model
matrices}\label{from-formula-to-model-matrices}

The process of converting a mixed-model formula for a data set into the
random-effects model matrix, \(\mbfZ\), the Gram matrix,
\(\mbfZ^{\top}\mbfZ\), and the mapping from the relative covariance
parameter, \(\mbftheta\), to the relative covariance factor,
\(\mbfLambda_\theta\), is described in considerable detail in
\citet[Sections 2.2 and 2.3]{bates.maechler.etal:2015}. In particular,
Tables 3 and 4 of \citet{bates.maechler.etal:2015} provide the
dimensions and the names of some of the components of these matrices and
should be kept close at hand when reading this section. For convenience,
Table 3 from \citet{bates.maechler.etal:2015} has been reproduced with
slight changes as Table~\ref{tbl-re-sizes} in Appendix \ref{sec-re-dim}.

To recap some of the material in \citet[Sections 2.2 and
2.3]{bates.maechler.etal:2015}, in a model with \(k\) random-effects
terms, an \(n\times p_i\) model matrix \(\mbfX_i, i=1,\dots,k\) is
evaluated from the expression \texttt{REterm}\(_i\) and the data. The
\(\mbfX_i, i=1,\dots,k\) are (implicitly) combined with the indicator
matrices, \(\mathbf{J}_i, i=1,\dots,k\) of the \(i\)th grouping factor
to produce the sparse model matrix blocks \(\mbfZ_i, i=1,\dots,k\). If
\(\ell_i, i = 1,\dots,k\) is the number of levels of the \(i\)th
grouping factor then the total number of the random effects for each
grouping factor is \(q_i=\ell_i p_i, i=1,\dots,k\) with
\(q=\sum_{i=1}^k q_i\) being the total number of random effects in the
model.

Generally the \(p_i, i=1,\dots,k\) are small. Random effects terms of
the form \texttt{(1\textbar{}g)}, i.e.~\(p_i=1\), are not uncommon. In
such cases, called simple, scalar random-effects terms, the matrix
\(\mbfZ_i\) is exactly the indicator matrix \(\mathbf{J}_i\) and its
\(\ell_i\times\ell_i\) Gram matrix, \(\mbfZ_i^{\top}\mbfZ_i\), is
diagonal with the incidence counts of the levels on the diagonal.

The blocked derivation in \citet{bates.maechler.etal:2015} carries over
into the model representation in \pkg{MixedModels.jl} with one
modification --- in \pkg{lme4} the (implicit) partitioning of \(\mbfZ\)
into \(k\) vertical blocks is according to the random-effects terms in
the formula, whereas in \pkg{MixedModels.jl} it is according to the
grouping factors. If more than one random-effects terms have the same
grouping factor they are amalgamated in the model representation.

Making the number of blocks small reduces the complexity of the
evaluation of the profiled likelihood while retaining simplicity in the
factorization of the first column of blocks (see
Section~\ref{sec-one-one-block}).

\subsection{An example using instructor evaluation
data}\label{an-example-using-instructor-evaluation-data}

The \texttt{InstEval} dataset from the \pkg{lme4} package is reproduced
as \texttt{insteval} in \pkg{MixedModels.jl}. (The \proglang{Julia}
convention is to use lower-case for names of objects and camel-case for
names of types and for package names.) It consists of nearly 75,000
evaluations (\texttt{y}) of instructors (\texttt{d} for \emph{Dozent},
German for ``instructor'') at ETH-Zürich by students (\texttt{s}). Other
covariates, such as the department (\texttt{dept}) that offered the
course and whether or not it was a \texttt{service} course, are
included. Further details are given on the \proglang{R} help page for
\texttt{lme4::InstEval}.

In \proglang{Julia} we attach the packages to be used, load the data
set, convert it to a \texttt{DataFrame} and evaluate some descriptive
statistics on its columns to produce Table~\ref{tbl-insteval-summary},
which shows that a total of 2972 students evaluated a total of 1128
instructors in courses from 14 different departments. About 43\% of the
evaluations were from service courses.

\begin{Shaded}
\begin{Highlighting}[]
\ImportTok{using} \BuiltInTok{AlgebraOfGraphics}\NormalTok{, }\BuiltInTok{CairoMakie}\NormalTok{, }\BuiltInTok{PrettyTables}\NormalTok{, }\BuiltInTok{Printf}\NormalTok{, }\BuiltInTok{SparseArrays}\NormalTok{, }\BuiltInTok{LinearAlgebra}
\ImportTok{using} \BuiltInTok{Arrow}\NormalTok{, }\BuiltInTok{Chairmarks}\NormalTok{, }\BuiltInTok{CSV}\NormalTok{, }\BuiltInTok{DataFrames}\NormalTok{, }\BuiltInTok{MixedModels}
\NormalTok{dat }\OperatorTok{=} \FunctionTok{DataFrame}\NormalTok{(MixedModels.}\FunctionTok{dataset}\NormalTok{(}\StringTok{"insteval"}\NormalTok{))}
\NormalTok{dat.service }\OperatorTok{=} \FunctionTok{float}\NormalTok{.(dat.service }\OperatorTok{.==} \StringTok{"Y"}\NormalTok{)       }\CommentTok{\# convert to 0/1 encoding}
\NormalTok{sumry }\OperatorTok{=} \FunctionTok{describe}\NormalTok{(dat, }\OperatorTok{:}\NormalTok{min, }\OperatorTok{:}\NormalTok{max, }\OperatorTok{:}\NormalTok{mean, }\OperatorTok{:}\NormalTok{nunique, }\OperatorTok{:}\NormalTok{eltype)}
\end{Highlighting}
\end{Shaded}

\begin{table}

\caption{\label{tbl-insteval-summary}Summary of the \texttt{insteval}
dataset}

\centering{

\begin{tabular}{rrrrrr}
  \hline
  \textbf{variable} & \textbf{min} & \textbf{max} & \textbf{mean} & \textbf{nunique} & \textbf{eltype} \\\hline
  s & S0001 & S2972 & - & 2972 & String \\
  d & I0001 & I2160 & - & 1128 & String \\
  dept & D01 & D15 & - & 14 & String \\
  studage & 2 & 8 & - & 4 & String \\
  lectage & 1 & 6 & - & 6 & String \\
  service & 0.0 & 1.0 & 0.43 & - & Float64 \\
  y & 1 & 5 & 3.21 & - & Int8 \\\hline
\end{tabular}

}

\end{table}%

We will use random effects to model the variation due to instructors and
due to students, as we regard the individual levels of these factors as
being samples from their respective populations. We will also use random
effects to model the variation due to departments and due to possible
department-to-department variation in the service-course effect.
(Technically this study is more of a census of departments than a random
sample but random effects are often used when the number of levels of
the factor is moderate, to help stabilize the computation.)

The model formula is chosen to illustrate certain aspects of the
conversion from formula to representation. In particular, we choose
independent random effects for \texttt{dept} and for \texttt{service} by
\texttt{dept} to show that distinct terms with the same grouping factor
will be amalgamated in the internal representation.

\begin{Shaded}
\begin{Highlighting}[]
\NormalTok{form }\OperatorTok{=} \PreprocessorTok{@formula}\NormalTok{(y }\OperatorTok{\textasciitilde{}} \FloatTok{1} \OperatorTok{+}\NormalTok{ service }\OperatorTok{+}\NormalTok{ (}\FloatTok{1}\OperatorTok{|}\NormalTok{d) }\OperatorTok{+}\NormalTok{ (}\FloatTok{1}\OperatorTok{|}\NormalTok{s) }\OperatorTok{+}\NormalTok{ (}\FloatTok{1}\OperatorTok{|}\NormalTok{dept) }\OperatorTok{+}\NormalTok{ (}\FloatTok{0} \OperatorTok{+}\NormalTok{ service}\OperatorTok{|}\NormalTok{dept))}
\NormalTok{m1 }\OperatorTok{=} \FunctionTok{fit}\NormalTok{(MixedModel, form, dat, progress}\OperatorTok{=}\ConstantTok{false}\NormalTok{) }\CommentTok{\# suppress display of a progress bar}
\FunctionTok{print}\NormalTok{(m1)}
\end{Highlighting}
\end{Shaded}

\begin{verbatim}
Linear mixed model fit by maximum likelihood
 y ~ 1 + service + (1 | d) + (1 | s) + (1 | dept) + (0 + service | dept)
    logLik     -2 logLik       AIC         AICc          BIC     
 -118824.3008  237648.6016  237662.6016  237662.6032  237727.0294

Variance components:
            Column    Variance  Std.Dev.   Corr.
s        (Intercept)  0.1052958 0.3244931
d        (Intercept)  0.2624286 0.5122778
dept     (Intercept)  0.0025800 0.0507934
         service      0.0233987 0.1529662   .  
Residual              1.3850086 1.1768639
 Number of obs: 73421; levels of grouping factors: 2972, 1128, 14

  Fixed-effects parameters:
─────────────────────────────────────────────────────
                  Coef.  Std. Error       z  Pr(>|z|)
─────────────────────────────────────────────────────
(Intercept)   3.27765     0.0235032  139.46    <1e-99
service      -0.0507433   0.0439878   -1.15    0.2487
─────────────────────────────────────────────────────
\end{verbatim}

Notice that, in the formula, the random effects term for instructor,
\texttt{(1\textbar{}d)}, occurred before that for student,
\texttt{(1\textbar{}s)}. In the model the \texttt{s} term occurs first
because the number of random effects from that term, 2972, is greater
than that from \texttt{d}, 1128. Also the two terms for \texttt{dept}
are amalgamated into a single term in the model.

The optimization of the profiled objective required about 175
evaluations of the objective function and took about a second on a
laptop computer, as shown in this benchmark using the \texttt{@be} macro
from the \pkg{Chairmarks.jl} package.

\begin{Shaded}
\begin{Highlighting}[]
\PreprocessorTok{@be} \FunctionTok{fit}\NormalTok{(}\OperatorTok{$}\NormalTok{MixedModel, }\OperatorTok{$}\NormalTok{form, }\OperatorTok{$}\NormalTok{dat; progress}\OperatorTok{=}\ConstantTok{false}\NormalTok{)  seconds}\OperatorTok{=}\FloatTok{8}
\end{Highlighting}
\end{Shaded}

\begin{verbatim}
Benchmark: 10 samples with 1 evaluation
 min    799.048 ms (233796 allocs: 62.013 MiB)
 median 803.584 ms (233796 allocs: 62.013 MiB, 0.69% gc time)
 mean   816.290 ms (233796 allocs: 62.013 MiB, 1.96% gc time)
 max    923.343 ms (233796 allocs: 62.013 MiB, 13.95% gc time)
\end{verbatim}

The environment in which these benchmarks were performed is described in
Section~\ref{sec-app-env}. Note that we use the Apple Accelerate
implementation of the BLAS/Lapack routines on computers with Apple
M-series processors and Intel's MKL BLAS implementation on computers
with x86\_64 processors. In Julia hot-swapping BLAS implementations is
as easy as loading the appropriate package, such as
\pkg{AppleAccelerate.jl} or \pkg{MKL.jl}.

Individual evaluations of the profiled objective, including installing a
new covariance parameter vector, updating the blocked Cholesky
factorization, and evaluating Equation~\ref{eq-profiled-log-likelihood},
takes a few milliseconds on the same laptop computer.

\begin{Shaded}
\begin{Highlighting}[]
\PreprocessorTok{@be} \FunctionTok{objective}\NormalTok{(}\FunctionTok{updateL!}\NormalTok{(}\FunctionTok{setθ!}\NormalTok{(}\OperatorTok{$}\NormalTok{m1, }\OperatorTok{$}\NormalTok{(m1.θ))))}
\end{Highlighting}
\end{Shaded}

\begin{verbatim}
Benchmark: 22 samples with 1 evaluation
 min    4.536 ms (58 allocs: 1.469 KiB)
 median 4.578 ms (58 allocs: 1.469 KiB)
 mean   4.579 ms (58 allocs: 1.469 KiB)
 max    4.652 ms (58 allocs: 1.469 KiB)
\end{verbatim}

The random effects terms define scalar random effects for \texttt{s} and
\texttt{d} and vector-valued random effects for \texttt{dept} in this
model, as reflected in the data type of each of the \texttt{reterms} of
the model

\begin{Shaded}
\begin{Highlighting}[]
\FunctionTok{show}\NormalTok{([}\FunctionTok{typeof}\NormalTok{(rt) for rt }\KeywordTok{in}\NormalTok{ m1.reterms])}
\end{Highlighting}
\end{Shaded}

\begin{verbatim}
DataType[ReMat{Float64, 1}, ReMat{Float64, 1}, ReMat{Float64, 2}]
\end{verbatim}

In the notation of \citet[Section 2]{bates.maechler.etal:2015} the
second parameter of each of these types is \(p_i, i = 1,\dots,3\).

The number of levels of the grouping factors are

\begin{Shaded}
\begin{Highlighting}[]
\FunctionTok{show}\NormalTok{([}\FunctionTok{length}\NormalTok{(rt.levels) for rt }\KeywordTok{in}\NormalTok{ m1.reterms])}
\end{Highlighting}
\end{Shaded}

\begin{verbatim}
[2972, 1128, 14]
\end{verbatim}

In the notation of \citet[Section 2]{bates.maechler.etal:2015} these are
\(\ell_i, i=1,\dots,3\).

Finally the number of columns in each of the \(\mbfZ_i, i=1,\dots,3\) is

\begin{Shaded}
\begin{Highlighting}[]
\FunctionTok{show}\NormalTok{([}\FunctionTok{size}\NormalTok{(rt, }\FloatTok{2}\NormalTok{) for rt }\KeywordTok{in}\NormalTok{ m1.reterms])}
\end{Highlighting}
\end{Shaded}

\begin{verbatim}
[2972, 1128, 28]
\end{verbatim}

For a summary of the model component sizes see Table~\ref{tbl-re-sizes}.

The blocks of the augmented Gram matrix, \(\mbfA\), and the lower
Cholesky factor, \(\mbfL_\theta=\mbfR^{\top}_\theta\), are summarized as

\begin{Shaded}
\begin{Highlighting}[]
\FunctionTok{BlockDescription}\NormalTok{(m1)}
\end{Highlighting}
\end{Shaded}

\begin{longtable}[]{@{}lllll@{}}
\caption{Block Structure of the \texttt{A} matrix
\label{tab:blk-descrip}}\tabularnewline
\toprule\noalign{}
rows & s & d & dept & fixed \\
\midrule\noalign{}
\endfirsthead
\toprule\noalign{}
rows & s & d & dept & fixed \\
\midrule\noalign{}
\endhead
\bottomrule\noalign{}
\endlastfoot
2972 & Diagonal & & & \\
1128 & Sparse & Diag/Dense & & \\
28 & Dense & Sparse/Dense & BlkDiag/Dense & \\
3 & Dense & Dense & Dense & Dense \\
\end{longtable}

(Only the lower triangle of the symmetric matrix, \(\mbfA\), is stored.)

In the theoretical development the blocking structure of the Gram
matrix, \(\mbfA\), and the Cholesky factor
(Equation~\ref{eq-bigCholfac}) is according to random effects, fixed
effects and response. In practice the response vector, \(\mbfy\), is
concatenated (\texttt{hcat}) to the fixed effects model matrix,
\(\mbfX\), and the random effects model matrix, \(\mbfZ\), is
partitioned according to the grouping factors.

In the block structure summary if a block of \(\mbfL\) and its
corresponding block of \(\mbfA\) have the same structure, the name of
that structure is given once. This is the case for all of the blocks
except the {[}2,2{]} block, which is diagonal in \(\mbfA\) but dense in
\(\mbfL\), the {[}2,3{]} block, which is sparse in \(\mbfA\) but dense
in \(\mbfL\), and the {[}3,3{]} block, which is block-diagonal in
\(\mbfA\) and dense in \(\mbfL\).

As described in \citet[Section 3]{bates.maechler.etal:2015} the relative
covariance factor, \(\mbfLambda_\theta\) is block diagonal with the same
blocking structure as the random effects parts of \(\mbfA\) or
\(\mbfL\). The \(i\)th diagonal block, \(i=1,\dots,k\), consists of
\(q_i\) repetitions on the diagonal of a \(\ell_i\times\ell_i\)
lower-triangular template matrix \(\mbfT_i\).

For a scalar random effects term (i.e.~\(\ell_i=1\)) its block in
\(\mbfLambda_\theta\) is a multiple of the \(q_i\times q_i\) identity
matrix. The blocks of \(\mbfLambda_\theta\) are never instantiated in
the \pkg{MixedModels.jl} implementation. There is sufficient flexibility
in the multiple dispatch model in \proglang{Julia} that
pre-multiplication by \(\mbfLambda_\theta^\top\) or post-multiplication
by \(\mbfLambda_\theta\) can be performed from the template matrices
\(\mbfT_i, i=1,\dots,k\) and the \texttt{refs} property of the grouping
factor alone.

The value of \(\hat{\mbftheta}\) for model \texttt{m1} is

\begin{Shaded}
\begin{Highlighting}[]
\NormalTok{θ̂ }\OperatorTok{=}\NormalTok{ m1.θ}
\FunctionTok{show}\NormalTok{(θ̂)}
\end{Highlighting}
\end{Shaded}

\begin{verbatim}
[0.2757269709081104, 0.4352906455775487, 0.04315999320792337, 0.12997785126273184]
\end{verbatim}

which maps to the template matrices

\begin{Shaded}
\begin{Highlighting}[]
\NormalTok{m1.λ}
\end{Highlighting}
\end{Shaded}

\begin{verbatim}
3-element Vector{AbstractMatrix{Float64}}:
 [0.2757269709081104;;]
 [0.4352906455775487;;]
 [0.04315999320792337 0.0; 0.0 0.12997785126273184]
\end{verbatim}

\subsection{Special role of the {[}1,1{]}
block}\label{sec-one-one-block}

The reason for ordering the random-effects terms so that the {[}1,1{]}
block of \(\mbfA\) is as large as possible is because the diagonal or
block diagonal structure of this block is preserved in the {[}1,1{]}
block of \(\mbfL\). To illustrate the importance of choosing the first
block of the augmented matrix \(\mbfA\), we fit the same model with (a)
the {[}1,1{]} block corresponding to the random effect term for the
instructor (\texttt{1\ \textbar{}\ d}) with 1128 levels, and (b) the
{[}1,1{]} block corresponding to the random effect term for students
(\texttt{(1\ \textbar{}\ s)} with 2972 levels).

\begin{figure}

\centering{

\includegraphics[width=6in,height=3.5in]{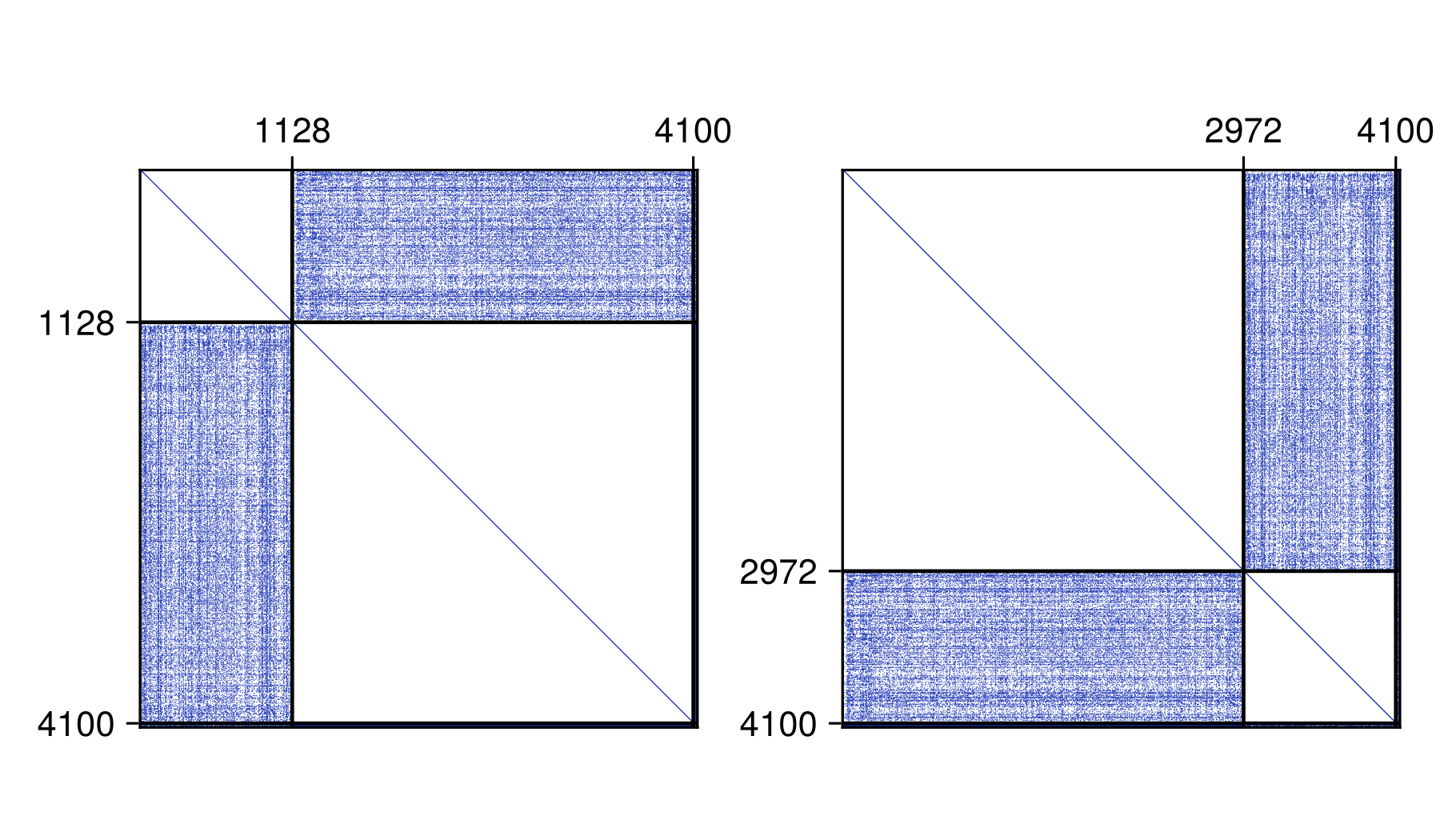}

}

\caption{\label{fig-a}Sparsity pattern of the augmented matrix \(\mbfA\)
when the order of the random effects is not carefully chosen (left) and
when the order is carefully chosen (right).}

\end{figure}%

\begin{figure}

\centering{

\includegraphics[width=6in,height=3.5in]{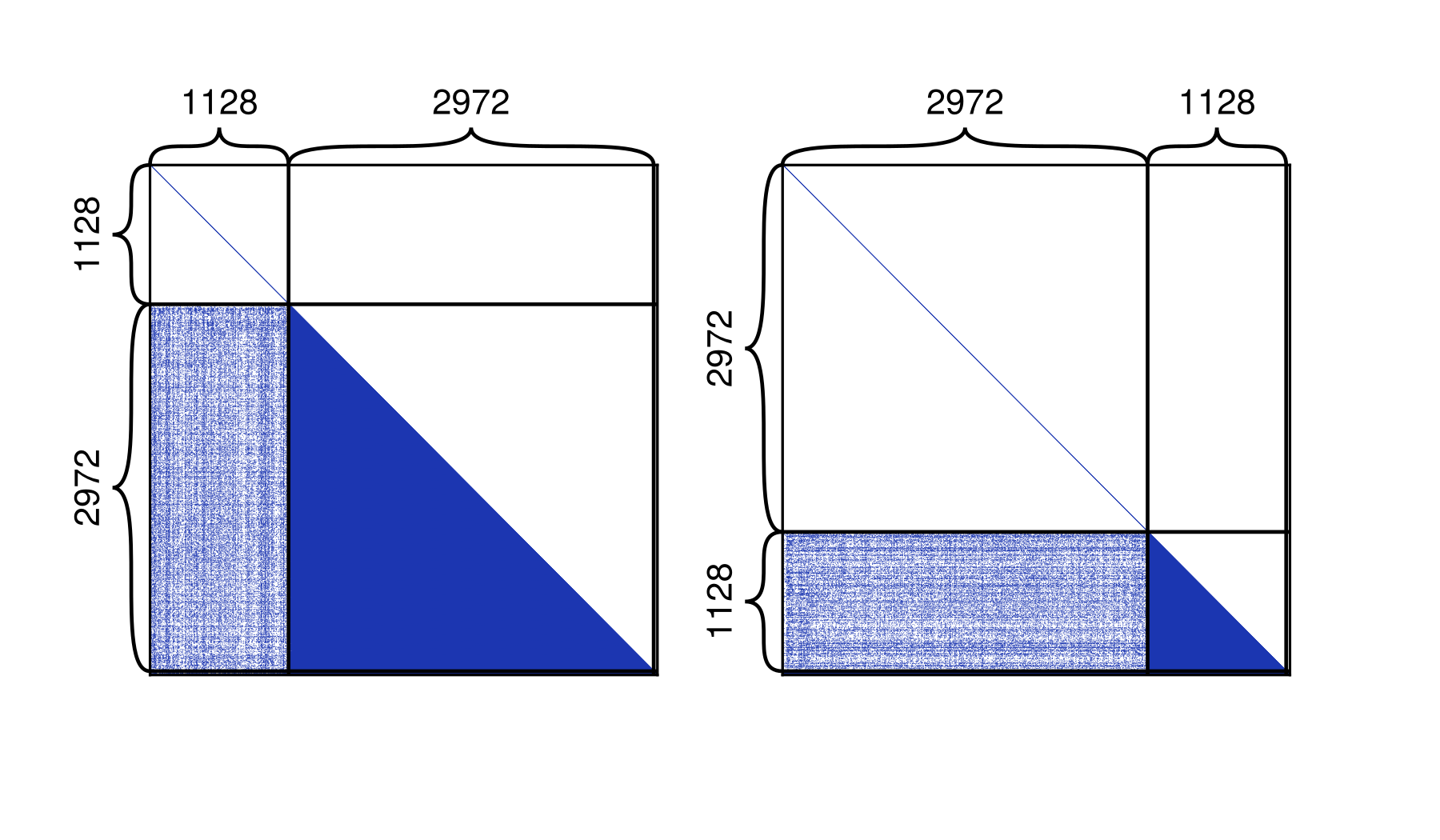}

}

\caption{\label{fig-blk}Sparsity pattern of the lower triangular
Cholesky factor when the order of the random effects is not carefully
chosen (left) and when the order is carefully chosen (right).}

\end{figure}%

Figure~\ref{fig-a} shows the non-zero elements of the augmented matrix,
\(\mbfA\), when (a) the first block corresponsds to the random effect
terms for the instructor (\texttt{(1\ \textbar{}\ d)} with 1128 levels),
and (b) the first block corresponds to the student
(\texttt{(1\ \textbar{}\ s)} with 2972 levels), and Figure~\ref{fig-blk}
shows the non-zero elements for the corresponding lower-triangular
Cholesky factor. When performing a block-wise Cholesky factorization,
the diagonal structure of the \([1,1]\) block from the input matrix is
preserved in the Cholesky factor, but the diagonal structure for the
subsequent diagonal blocks (e.g {[}2,2{]}) is not preserved. While the
models are theoretically identical, the latter is computationally more
efficient as the largest diagonal block preserves its structure. In
fact, for the \texttt{insteval} dataset, the former order leads to
approximately 4.5 million non-zero entries in the Cholesky factor,
whereas choosing the random effect with the largest number of levels
leads to a Cholesky factor with approximately 775,000 non-zero elements,
which is an order of magnitude lower. This translates to an order of
magnitude difference between the runtimes for fitting the two versions
of the model. We demonstrate this with the \texttt{insteval} dataset in
Appendix \ref{sec-app-re} where fitting the model with the two blocks
swapped is almost twenty times slower.

Because the {[}1,1{]} block of \(\mbfL\) is diagonal and the {[}1,1{]}
block of \(\mbfLambda_\theta\) is a multiple of the identity, the
remaining blocks in the first column of blocks of \(\mbfL\) will have
exactly the same sparsity structure as the corresponding blocks of
\(\mbfA\). For example, the {[}2,1{]} block of \(\mbfA\) is stored as a
sparse matrix because only about 2\% (73421) of the
\(2792\times1128=3352416\) elements of that matrix are non-zero. The
{[}2,1{]} block of \(\mbfL\) has exactly the same sparsity structure
because each element of that block of \(\mbfL\) is a scalar multiple of
the corresponding element of that block of \(\mbfA\).

\section{Application to data from an observational
study}\label{sec-application}

We consider an observational study -- ratings of movies by users of
\href{https://movielens.org}{movielens.org} made available at the
\href{https://grouplens.org/datasets/movielens}{grouplens.org download
site} \citep{Harper2016} as \emph{MovieLens 32M Dataset}
(\texttt{ml-32m}), consisting of roughly 32 million ratings of over
80,000 movies by over 200,000 users. The purpose of this section is to
study which dimensions of the data have the greatest effect on the
amount of memory used to represent the model and the time required to
fit a model. To do so we use a simple model:

\begin{Shaded}
\begin{Highlighting}[]
\NormalTok{frm }\OperatorTok{=} \PreprocessorTok{@formula}\NormalTok{(rating }\OperatorTok{\textasciitilde{}} \FloatTok{1} \OperatorTok{+}\NormalTok{ (}\FloatTok{1} \OperatorTok{|}\NormalTok{ userId) }\OperatorTok{+}\NormalTok{ (}\FloatTok{1} \OperatorTok{|}\NormalTok{ movieId))}
\FunctionTok{fit}\NormalTok{(LinearMixedModel, frm, dataset)}
\end{Highlighting}
\end{Shaded}

where \texttt{dataset} will be different subsets of the full data
explained below.

\subsection{Structure of the data}\label{structure-of-the-data}

The \texttt{ratings.csv} file in the \texttt{ml-32m.zip} file downloaded
from the Grouplens site provides a table of about 32 million ratings
with the corresponding movie number and user number. Creating a
\texttt{ratings} data frame from this file and counting the number of
ratings for each movie id provides the \texttt{movies} data frame with
columns \texttt{movieId} and \texttt{mnrtngs}. Similarly counting the
number of ratings for each user provides the \texttt{users} data frame
with columns \texttt{userId} and \texttt{unrtngs}. Joining these two
derived data frames with the original \texttt{ratings} data frame (on
\texttt{movieId} and \texttt{userId}, respectively) allows for the
\texttt{ratings} data frame to be subsetted according to a minimum
number of ratings per user and/or per movie.

\begin{figure}

\centering{

\pandocbounded{\includegraphics[keepaspectratio]{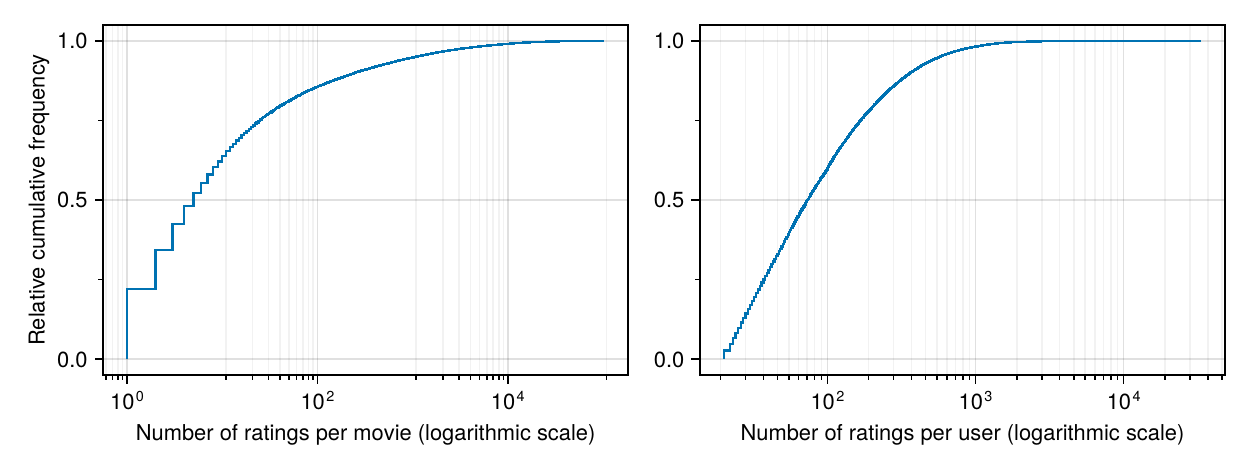}}

}

\caption{\label{fig-nrtngsecdf}Empirical distribution plots of the
number of ratings per movie and per user. The horizontal axes are on a
logarithmic scale.}

\end{figure}

The data from this observational study are extremely unbalanced with
respect to the grouping factors, \texttt{userId} and \texttt{movieId},
as shown in the empirical cumulative distribution function (ecdf) plots
in Figure~\ref{fig-nrtngsecdf}. (Note that the horizontal axis in each
panel is on a logarithmic scale.)

This selection of ratings was limited to users who had provided at least
20 ratings.

In this collection of about 32 million ratings, over 22\% of the movies
are rated only once. The median number of ratings is 5 but the maximum
is about 103,000.

The ecdf plot of the number of ratings per user shows a similar pattern
to that of the movies --- a few users with a very large number of
ratings and many users with just a few ratings.

For example, the minimum number of movies rated is 20 (due to the
inclusion constraint); the median number of movies rated is around 70;
but the maximum is over 33,000 (which is a lot of movies--over 30 years
of 3 movies per day every day--if this user actually watched all of the
movies they rated).

Movies with very few ratings provide little information about overall
trends or even about the movie being rated. We can imagine that the
``shrinkage'' of random effects for movies with just a few ratings pulls
their adjusted rating strongly towards the overall average.

Similarly, users who rate very few movies add little information, even
about the movies that they rated, because there isn't sufficient
information to contrast a specific rating with a typical rating for the
user.

One way of dealing with the extreme imbalance in the number of
observations per user or per movie is to set a minimum number of ratings
for a user or for a movie to be included in the data used to fit the
model.

\subsection{Models fit with lower bounds on ratings per user and per
movie}\label{sec-lrgobsmods}

We fit a simple model to this dataset using different thresholds on the
number of ratings per movie and the number of ratings per user. These
fits were performed on a desktop computer with a generous amount of
memory (96 GiB) and using 8 threads and the Intel Math Kernel Library
(MKL) for numerical linear algebra. The results are summarized in
Table~\ref{tbl-sizespeed} in Appendix \ref{sec-app-ml}.

\subsubsection{Dimensions of the model versus cut-off
values}\label{dimensions-of-the-model-versus-cut-off-values}

\begin{figure}

\centering{

\pandocbounded{\includegraphics[keepaspectratio]{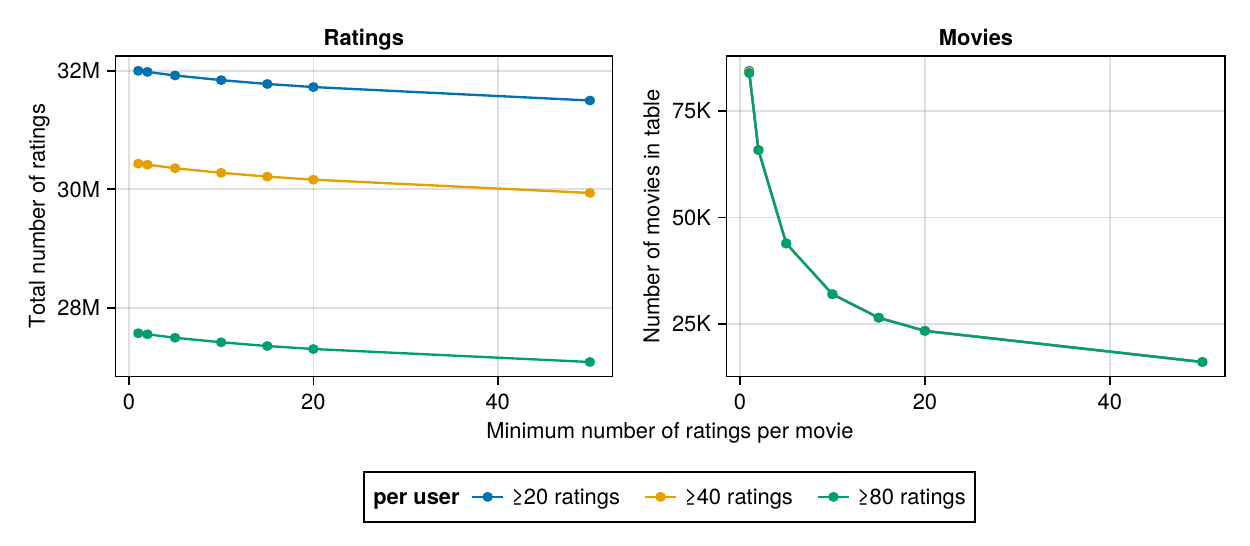}}

}

\caption{\label{fig-nratingsbycutoff}Number of ratings (left) and movies
(right) in reduced table by movie cutoff and by user cutoff}

\end{figure}

As shown in the first panel of Figure~\ref{fig-nratingsbycutoff}, the
number of ratings varies from a little over 21 million to 32 million.
For this set of cutoff values, the user cutoff has more impact on the
number of ratings in the reduced dataset than does the movie cutoff.

A glance at Table~\ref{tbl-sizespeed} shows that the number of users is
essentially a function of only the user cutoff.

The second panel in Figure~\ref{fig-nratingsbycutoff} shows the
similarly unsurprising result that the number of movies in the reduced
table is essentially determined by the movie cutoff.

\subsubsection{Memory footprint of the model
representation}\label{sec-lrgobsmemprint}

\begin{figure}

\centering{

\pandocbounded{\includegraphics[keepaspectratio]{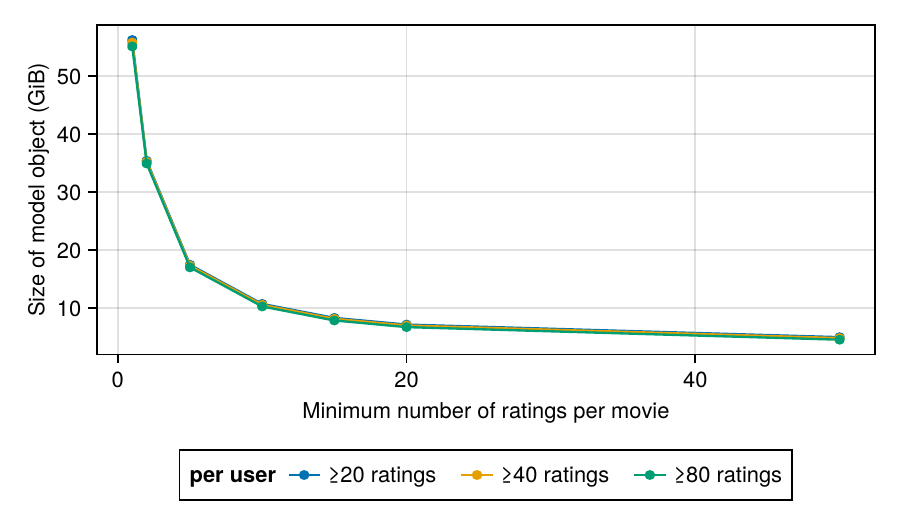}}

}

\caption{\label{fig-memoryfootprint}Memory footprint of the model
representation by minimum number of ratings per user and per movie.}

\end{figure}

Figure~\ref{fig-memoryfootprint} shows that the overall memory size (the
``memory footprint'') of the model representation depends primarily on
the movie cutoff. (There are three lines on
Figure~\ref{fig-memoryfootprint}, corresponding to the three user cutoff
values, but they are nearly coincident.)

\begin{figure}

\centering{

\pandocbounded{\includegraphics[keepaspectratio]{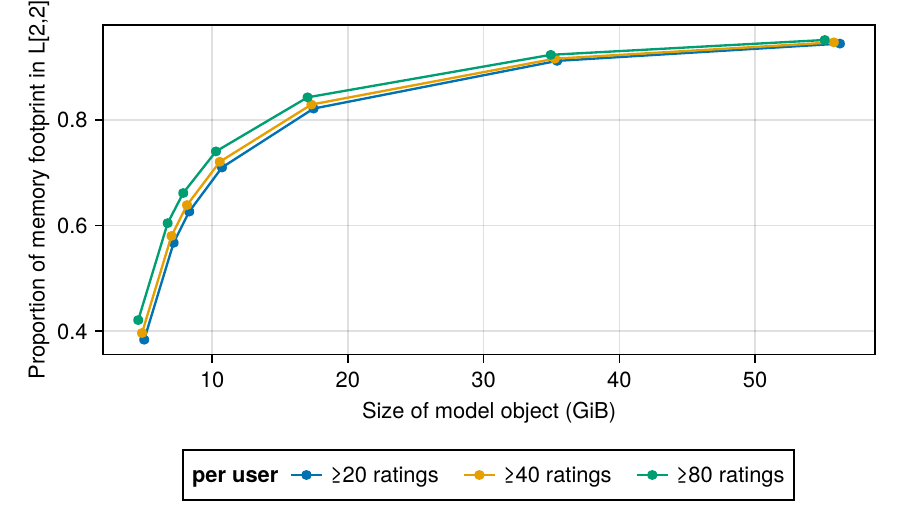}}

}

\caption{\label{fig-l22prop}Proportion of memory footprint of the model
in L{[}2,2{]} versus the overall model size (GiB).}

\end{figure}

Figure~\ref{fig-l22prop} shows the dominance of the \texttt{L{[}2,2{]}}
block, which is a dense lower triangular matrix, in the overall memory
footprint of the model. When all the movies are included in the data to
which the model is fit the total memory footprint is over 50 GiB, and
over 90\% of that memory is for \texttt{L{[}2,2{]}}. Even when requiring
a minimum of 80 ratings per movie, \texttt{L{[}2,2{]}} accounts for
about 40\% of the memory footprint.

The fact that the \texttt{{[}2,2{]}} block of \texttt{A} is diagonal but
the \texttt{{[}2,2{]}} block of \texttt{L} is dense lower-triangular is
described as ``fill-in'' and leads to a somewhat unintuitive conclusion.
The memory footprint of the model representation depends strongly on the
number of movies, less strongly on the number of users and almost not at
all on the number of ratings (Figure~\ref{fig-memoryfootprint}).

Note that although the dimensions of the \texttt{L{[}2,1{]}} block are
larger than those of the \texttt{L{[}2,2{]}} block its memory footprint
is smaller than that of the \texttt{{[}2,2{]}} block, because it is
stored as a sparse matrix. The matrix is about 98\% zeros or,
equivalently, a little over 2\% nonzeros, which makes the sparse
representation much smaller than the dense representation. Also, the
number and positions of the non-zeros in \texttt{L{[}2,1{]}} is exactly
the same as those of in \texttt{A{[}2,1{]}}. That is, there is no
fill-in in the \texttt{{[}2,1{]}} block

In a sense this is good news because the amount of storage required for
the \texttt{{[}2,2{]}} block can be nearly cut in half by taking
advantage of the fact that it is a triangular matrix. See
Section~\ref{sec-RFP} for further discussion on one approach using the
\emph{rectangular full packed format} \citep{lawn199}.

In general, for models with scalar random effects for two incompletely
crossed grouping factors, the memory footprint depends strongly on the
smaller of the number of levels of the grouping factors, less strongly
on the larger number of levels, and almost not at all on the number of
observations.

\subsubsection{Speed of log-likelihood
evaluation}\label{speed-of-log-likelihood-evaluation}

The time required to fit a model to large data sets is dominated by the
time required to evaluate the log-likelihood during the optimization of
the parameter estimates. The time for one evaluation is given in the
\texttt{time\ per\ eval} column of Table~\ref{tbl-sizespeed}. Also given
is the number of evaluations to convergence, \texttt{n\ eval}, and the
total time to fit the model, \texttt{time}. The reason for considering
\texttt{evtime} in addition to \texttt{fittime} and \texttt{n\ eval} is
because the \texttt{time\ per\ eval} for one model, relative to other
models, is reasonably stable across computers whereas \texttt{n\ eval},
and hence, \texttt{time}, can be affected by seemingly trivial
variations in function values resulting from different implementations
of low-level calculations, such as the BLAS (Basic Linear Algebra
Subroutines). (Floating point arithmetic is not guaranteed to be
associative and custom BLAS implementations may rearrange the order of
operations when evaluating products and sums.)

That is, we can't expect to reproduce \texttt{n\ eval} exactly when
fitting the same model on different computers or with slightly different
versions of software; but we can expect the pattern in
\texttt{time\ per\ eval} with respect to the user and movie cutoffs to
be reproducible.

\begin{figure}

\centering{

\pandocbounded{\includegraphics[keepaspectratio]{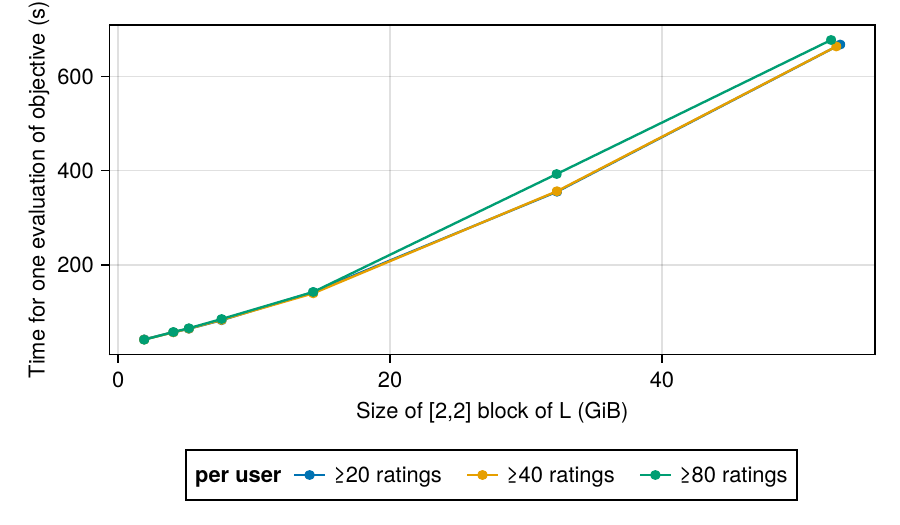}}

}

\caption{\label{fig-evtimevsl22}Evaluation time for the objective (s)
versus size of the {[}2,2{]} block of L (GiB)}

\end{figure}

As shown in Figure~\ref{fig-evtimevsl22} the evaluation time for the
objective is predominantly a function of the size of the
\texttt{{[}2,2{]}} block of \texttt{L}.

\subsubsection{Reducing the amount of storage required}\label{sec-RFP}

As shown in Figure~\ref{fig-l22prop} the \texttt{L{[}2,2{]}} block, when
stored as a dense square matrix, can be a major contributor to the
amount of storage required for the model. Because only the lower
triangle of this dense square matrix is used, about half of this storage
is never accessed. The rectangular full-packed (RFP) storage format
\citep{lawn199} allows for a triangular matrix to be packed into
(roughly) half the amount of storage required for the dense square
matrix. The \pkg{RectangularFullPacked.jl} package allows access to this
format in \proglang{Julia}.

The savings in storage comes at the expense of a more complicated
algorithm for mapping matrix coordinates to storage locations.

The LAPACK implementation of the Cholesky factorization for the RFP
format (\texttt{dpftrf}) is nearly as fast as the Cholesky factorization
for a dense square matrix (\texttt{dpotrf}). However, the blocked
Cholesky factorization requires several operations to evaluate the
matrix to be passed to LAPACK's dense Cholesky factorization. The most
expensive of these occurs after \(\mbfL[2,1]\) and the diagonal matrix
\(\mbfLambda_2^\top\mbfA[2,2]\mbfLambda_2+\mbfI\) have been evaluated
(in the storage for \(\mbfL[2,2]\)). It is described as a ``symmetric
rank-k update'' and involves replacing the current \(\mbfL[2,2]\) by
\(\mbfL[2,2]-\mbfL[2,1]\mbfL[2,1]^\top\).

In practical examples, \(\mbfL[2,1]\) is sparse and the update is
performed by iterating over columns of \(\mbfL[2,1]\). Each pair of
nonzeros in each column of \(\mbfL[2,1]\) generates a (scalar) update of
one of the elements of \(\mbfL[2,2]\) and this is where the more
expensive indexing system in the RFP formulation (relative to storing a
square dense matrix) takes its toll. Our current implementation using
RFP takes three to four times as long to perform the rank-k update than
to do the factorization to obtain \(\mbfL[2,2]\).

\subsubsection{A{[}2,1{]} as a biadjacency
matrix}\label{sec-biadjacency}

In both of the examples we have considered, the first and second
random-effects terms are simple scalar terms and each observation
corresponds to a unique combination of levels of these two factors. For
example, in the movielens data each rating corresponds to a unique
combination of a user and a movie.

In cases like these, the \(\mbfA[2,1]\) matrix is the biadjacency matrix
for a bipartite graph \((U,V,E)\) where the nodes \(U\) are the
individual users, the nodes \(V\) are the individual movies and the
edges, \(E\), link a movie to a user who rated it. The total number of
edges in the bipartite graph is \(n\), the number of observations in the
data set. Furthermore, the diagonal elements in the diagonal matrices
\(\mbfA[1,1]\) and \(\mbfA[2,2]\) are the degrees of the vertices \(U\)
and \(V\), respectively. They are also the column- and row-sums of
\(\mbfA[2,1]\). It may be possible to derive new computational
approaches from these relationships using graph theory, an avenue we
hope to explore in future research.

\section{Further Comments}\label{sec-further-comments}

Throughout the discussion we have considered maximum likelihood
estimation, which is the default estimation criterion. However, it is
also possible to fit mixed models using the restricted maximum
likelihood (REML) criterion by setting the optional argument
\texttt{REML=true} in the call to the \texttt{fit} function. Briefly,
the maximum likelihood estimates of the random effects variance
components are biased. The REML estimates provide less biased estimates
of variance components. For more details see \citep[see][Section
3.4]{bates.maechler.etal:2015}. For instance:

\begin{Shaded}
\begin{Highlighting}[]
\NormalTok{mreml }\OperatorTok{=} \FunctionTok{fit}\NormalTok{(}
\NormalTok{  MixedModel,}
  \PreprocessorTok{@formula}\NormalTok{(y }\OperatorTok{\textasciitilde{}} \FloatTok{1} \OperatorTok{+}\NormalTok{ service }\OperatorTok{+}\NormalTok{ (}\FloatTok{1} \OperatorTok{|}\NormalTok{ d) }\OperatorTok{+}\NormalTok{ (}\FloatTok{1} \OperatorTok{|}\NormalTok{ s) }\OperatorTok{+}\NormalTok{ (}\FloatTok{1} \OperatorTok{|}\NormalTok{ dept) }\OperatorTok{+}\NormalTok{ (}\FloatTok{0} \OperatorTok{+}\NormalTok{ service }\OperatorTok{|}\NormalTok{ dept)),}
\NormalTok{  dat,}
\NormalTok{  REML }\OperatorTok{=} \ConstantTok{true}\NormalTok{,}
\NormalTok{  progress}\OperatorTok{=}\ConstantTok{false}
\NormalTok{)}
\FunctionTok{print}\NormalTok{(mreml)}
\end{Highlighting}
\end{Shaded}

\begin{verbatim}
Linear mixed model fit by REML
 y ~ 1 + service + (1 | d) + (1 | s) + (1 | dept) + (0 + service | dept)
 REML criterion at convergence: 237658.60945245498

Variance components:
            Column    Variance  Std.Dev.   Corr.
s        (Intercept)  0.1053198 0.3245301
d        (Intercept)  0.2624398 0.5122888
dept     (Intercept)  0.0030492 0.0552198
         service      0.0256136 0.1600426   .  
Residual              1.3850023 1.1768612
 Number of obs: 73421; levels of grouping factors: 2972, 1128, 14

  Fixed-effects parameters:
─────────────────────────────────────────────────────
                  Coef.  Std. Error       z  Pr(>|z|)
─────────────────────────────────────────────────────
(Intercept)   3.27771     0.0242443  135.20    <1e-99
service      -0.0502837   0.0457707   -1.10    0.2719
─────────────────────────────────────────────────────
\end{verbatim}

The estimates obtained above are the same as those obtained from
\texttt{lme4::lmer} function in \proglang{R}.

We used the formula
\texttt{y\ \textasciitilde{}\ 1\ +\ service\ +\ (1\ \textbar{}\ d)\ +\ (1\ \textbar{}\ s)\ +\ (1\ \textbar{}\ dept)\ +\ (0\ +\ service\ \textbar{}\ dept)}
to demonstrate that \pkg{MixedModels.jl} amalgamates random effects
terms with the same grouping factor. In practice we can use the more
concise formula:
\texttt{y\ \textasciitilde{}\ 1\ +\ service\ +\ (1\ \textbar{}\ d)\ +\ (1\ \textbar{}\ s)\ +\ zerocorr(service\ \textbar{}\ dept)}.
That is, the term \texttt{zerocorr(service\ \textbar{}\ dept)} is
equivalent to adding the two terms
\texttt{(1\ \textbar{}\ dept)\ +\ (0\ +\ service\ \textbar{}\ dept)},
and the same as the \pkg{lme4} formula term
\texttt{(1\ +\ service\ \textbar{}\textbar{}\ dept)}.

Although we only demonstrated the fitting of linear mixed models,
\pkg{MixedModels.jl} can also fit generalized linear mixed models
including logistic, probit, poisson, and negative binomial models. To
fit a probit model for instance we use the following syntax:

\begin{Shaded}
\begin{Highlighting}[]
\NormalTok{probmod }\OperatorTok{=} \FunctionTok{fit}\NormalTok{(MixedModel, frm, dt, }\FunctionTok{Bernoulli}\NormalTok{(), }\FunctionTok{ProbitLink}\NormalTok{())}
\end{Highlighting}
\end{Shaded}

where \texttt{frm} is a formula and \texttt{dt} is a dataset. Refer to
the \pkg{MixedModels.jl} documentation for more details.

\section{Summary and discussion}\label{sec-summary}

We demonstrated how the \pkg{MixedModels.jl} package can be used to fit
models with two distinct grouping factors (the \texttt{ml-32m} example)
and three distinct grouping factors (\texttt{insteval} example). Fitting
models with a single grouping factor or with more than three grouping
factors is straightforward.

The key contribution of \pkg{MixedModels.jl} is the new fitting
procedure based on the augmented matrix \(\mbfA\) (Equation~\ref{eq-A})
and the simplified formula for the profiled loglikelihood
(Equation~\ref{eq-profiled-log-likelihood}). As a result of this
simplification, we do not need to compute the conditional mode of the
random effects (\(\tilde{\mbfu}\) or \(\tilde{\mbfb}\)) for each
evaluation of the objective function. This means we avoid roughly
\(O(n)\) computations at each evaluation as compared to previous
implementations of mixed effects models, making the algorithm more
scalable. In fact the conditional modes are computed only once at the
converged values of the underlying parameter \(\mbftheta\). Moreover, at
each iteration we only need to use the augmented Gram matrix \(\mbfA\)
(Equation~\ref{eq-A}), which is a blocked, symmetric, sparse matrix of
size \(q + p + 1\) and the covariance parameter vector, \(\mbftheta\),
to form the blocked Cholesky factor, \(\mbfL\). This update is performed
in-place.

Secondly, by rearranging the random effects terms so that the grouping
factor with the greatest number of random effects corresponds to the
\([1,1]\) block of the Cholesky factor, we reduce the ``fill-in'' of the
Cholesky factor. These two changes together contribute to a substantial
improvement in computational efficiency.

Finally, the multiple dispatch feature of the \proglang{Julia} language
allows us to use specialized algorithms for different kinds of matrices
instead of relying on generic algorithms. This is relevant because we
have both sparse and dense operations in different blocks of the
Cholesky factor. By dispatching on the type of the block, we can create
performant algorithms for each combination of matrix types that can
occur.

\section*{Acknowledgments}\label{acknowledgments}
\addcontentsline{toc}{section}{Acknowledgments}

This research was supported by the Center for Interdisciplinary
Research, Bielefeld (ZiF) Cooperation Group ``Statistical models for
psychological and linguistic data''. We would like to thank Ben Bolker
for helpful comments on the manuscript.

All figures were produced using the \pkg{AlgebraOfGraphics.jl} package
with the \pkg{CairoMakie.jl} backend in the \pkg{Makie.jl} graphics
system \citep{DanischKrumbiegel2021}. Julius Krumbiegel and Anshul
Singhvi provided invaluable help and feedback with \pkg{Makie.jl}. The
plots in this work would not have been possible without them. Finally,
we would like to thank the \proglang{Julia} community for bugfixes and
answering our questions in the community support channels.

\section*{References}\label{references}
\addcontentsline{toc}{section}{References}

\renewcommand{\bibsection}{}
\bibliography{ms}

\newpage{}

\appendix
\renewcommand{\thefigure}{A\arabic{figure}}
\renewcommand{\thetable}{A\arabic{table}}
\setcounter{figure}{0}
\setcounter{table}{0}

\section{Appendix}\label{appendix}

\subsection{Comparative benchmarks with lme4 and
glmmTMB}\label{comparative-benchmarks-with-lme4-and-glmmtmb}

We benchmarked fitting the same model as in Section~\ref{sec-structure}
using the \pkg{lme4} and \pkg{glmmTMB} packages for \proglang{R}.

\begin{verbatim}
> library(lme4, quietly = TRUE)
> library(glmmTMB)
> library(microbenchmark)
> dat <- InstEval
> dat$service <- as.numeric(dat$service) - 1  # convert to a 0/1 numeric vector
> ctrl <- lmerControl(calc.derivs = FALSE)
> form <- y ~ 1 + service + (1 | s) + (1 | d) + (1 | dept) + (0 + service | dept)
> mlmer <- lmer(form, dat, REML = FALSE, control = ctrl)
> summary(mlmer, correlation = FALSE)
Linear mixed model fit by maximum likelihood  ['lmerMod']
Formula: y ~ 1 + service + (1 | s) + (1 | d) + (1 | dept) + (0 + service |  
    dept)
   Data: dat
Control: ctrl

      AIC       BIC    logLik -2*log(L)  df.resid 
 237662.6  237727.0 -118824.3  237648.6     73414 

Scaled residuals: 
     Min       1Q   Median       3Q      Max 
-2.99959 -0.74769  0.04009  0.77283  3.11345 

Random effects:
 Groups   Name        Variance Std.Dev.
 s        (Intercept) 0.105295 0.32449 
 d        (Intercept) 0.262425 0.51227 
 dept     (Intercept) 0.002579 0.05079 
 dept.1   service     0.023405 0.15299 
 Residual             1.385009 1.17686 
Number of obs: 73421, groups:  s, 2972; d, 1128; dept, 14

Fixed effects:
            Estimate Std. Error t value
(Intercept)  3.27765    0.02350 139.463
service     -0.05074    0.04399  -1.153
> tmbmod <- glmmTMB(form, dat)
> summary(tmbmod)
 Family: gaussian  ( identity )
Formula:          
y ~ 1 + service + (1 | s) + (1 | d) + (1 | dept) + (0 + service |      dept)
Data: dat

      AIC       BIC    logLik -2*log(L)  df.resid 
 237662.6  237727.0 -118824.3  237648.6     73414 

Random effects:

Conditional model:
 Groups   Name        Variance Std.Dev.
 s        (Intercept) 0.10530  0.32449 
 d        (Intercept) 0.26243  0.51228 
 dept     (Intercept) 0.00258  0.05079 
 dept.1   service     0.02340  0.15296 
 Residual             1.38501  1.17686 
Number of obs: 73421, groups:  s, 2972; d, 1128; dept, 14

Dispersion estimate for gaussian family (sigma^2): 1.39 

Conditional model:
            Estimate Std. Error z value Pr(>|z|)    
(Intercept)  3.27765    0.02351  139.39   <2e-16 ***
service     -0.05074    0.04408   -1.15     0.25    
---
Signif. codes:  0 ‘***’ 0.001 ‘**’ 0.01 ‘*’ 0.05 ‘.’ 0.1 ‘ ’ 1
> microbenchmark(
+   lmer = lmer(form, dat, REML = FALSE, control = ctrl),
+   glmmTMB = glmmTMB(form, dat), times = 6
+ )
Unit: seconds
    expr      min       lq     mean   median       uq      max neval cld
    lmer 17.29331 17.32405 17.45185 17.44362 17.49485 17.71163     6  a 
 glmmTMB 22.60197 22.61996 22.65325 22.64604 22.65600 22.74950     6   b
> # R and package versions
> R.version.string
[1] "R version 4.5.0 (2025-04-11)"
> packageVersion("lme4")
[1] ‘1.1.37’
> packageVersion("glmmTMB")
[1] ‘1.1.11’
> 
\end{verbatim}

As of writing, an attempt to fit the \texttt{ml-32m} example using
\pkg{glmmTMB} has been running for more than a week on the machine that
was used for the \pkg{MixedModels.jl} timings and has not yet completed.
Given otherwise similar performance, we do not expect \pkg{lme4} to
perform any better than \pkg{glmmTMB}.

\subsection{Ordering of the random effects}\label{sec-app-re}

To demonstrate the effect of ordering of the random effects, we fit the
\texttt{insteval} model after a different ordering for the random
effects. We first recreate a copy of the \texttt{insteval} model

\begin{Shaded}
\begin{Highlighting}[]
\NormalTok{m2 }\OperatorTok{=} \FunctionTok{fit}\NormalTok{(}
\NormalTok{  MixedModel,}
  \PreprocessorTok{@formula}\NormalTok{(y }\OperatorTok{\textasciitilde{}} \FloatTok{1} \OperatorTok{+}\NormalTok{ service }\OperatorTok{+}\NormalTok{ (}\FloatTok{1} \OperatorTok{|}\NormalTok{ d) }\OperatorTok{+}\NormalTok{ (}\FloatTok{1} \OperatorTok{|}\NormalTok{ s) }\OperatorTok{+}\NormalTok{ (}\FloatTok{1} \OperatorTok{|}\NormalTok{ dept) }\OperatorTok{+}\NormalTok{ (}\FloatTok{0} \OperatorTok{+}\NormalTok{ service }\OperatorTok{|}\NormalTok{ dept)),}
\NormalTok{  dat,}
\NormalTok{  progress}\OperatorTok{=}\ConstantTok{false}    \CommentTok{\# suppress the display of a progress bar}
\NormalTok{);}
\end{Highlighting}
\end{Shaded}

To demonstrate, we must manually reorder the random effects then
recreate the augmented matrix \(\mbfA\) and storage for the Cholesky
factor \(\mbfL\) as follows:

\begin{Shaded}
\begin{Highlighting}[]
\NormalTok{m2.reterms[}\FloatTok{1}\OperatorTok{:}\FloatTok{2}\NormalTok{] }\OperatorTok{=}\NormalTok{ m2.reterms[}\FloatTok{2}\OperatorTok{:{-}}\FloatTok{1}\OperatorTok{:}\FloatTok{1}\NormalTok{]; }\CommentTok{\# swap elements 1 and 2}

\NormalTok{A, L }\OperatorTok{=}\NormalTok{ MixedModels.}\FunctionTok{createAL}\NormalTok{(m2.reterms, m2.Xymat); }\CommentTok{\# recreate A, L}

\FunctionTok{copyto!}\NormalTok{(m2.A, A); }\CommentTok{\# set A}
\FunctionTok{copyto!}\NormalTok{(m2.L, L); }\CommentTok{\# set L}
\end{Highlighting}
\end{Shaded}

Now we can refit the model:

\begin{Shaded}
\begin{Highlighting}[]
\PreprocessorTok{@be} \FunctionTok{refit!}\NormalTok{(}\OperatorTok{$}\NormalTok{m2; progress }\OperatorTok{=} \ConstantTok{false}\NormalTok{) seconds }\OperatorTok{=} \FloatTok{100} \CommentTok{\# very slow}
\end{Highlighting}
\end{Shaded}

\begin{verbatim}
Benchmark: 8 samples with 1 evaluation
 min    12.699 s (11711 allocs: 329.203 KiB)
 median 12.770 s (11711 allocs: 329.203 KiB)
 mean   12.767 s (11711 allocs: 329.203 KiB)
 max    12.855 s (11711 allocs: 329.203 KiB)
\end{verbatim}

This is an order of magnitude slower than just refitting the original
model \texttt{m1}:

\begin{Shaded}
\begin{Highlighting}[]
\PreprocessorTok{@be} \FunctionTok{refit!}\NormalTok{(}\OperatorTok{$}\NormalTok{m1, progress }\OperatorTok{=} \ConstantTok{false}\NormalTok{) seconds }\OperatorTok{=} \FloatTok{8}
\end{Highlighting}
\end{Shaded}

\begin{verbatim}
Benchmark: 11 samples with 1 evaluation
 min    772.174 ms (11239 allocs: 285.078 KiB)
 median 772.347 ms (11239 allocs: 285.078 KiB)
 mean   772.921 ms (11239 allocs: 285.078 KiB)
 max    777.291 ms (11239 allocs: 285.078 KiB)
\end{verbatim}

Note that the output is exactly the same as \texttt{m1} but it takes
much longer to fit.

\begin{Shaded}
\begin{Highlighting}[]
\FunctionTok{print}\NormalTok{(m2)}
\end{Highlighting}
\end{Shaded}

\begin{verbatim}
Linear mixed model fit by maximum likelihood
 y ~ 1 + service + (1 | d) + (1 | s) + (1 | dept) + (0 + service | dept)
    logLik     -2 logLik       AIC         AICc          BIC     
 -118824.3008  237648.6016  237662.6016  237662.6032  237727.0294

Variance components:
            Column    Variance  Std.Dev.   Corr.
d        (Intercept)  0.2624276 0.5122769
s        (Intercept)  0.1052962 0.3244938
dept     (Intercept)  0.0025796 0.0507898
         service      0.0233958 0.1529568   .  
Residual              1.3850086 1.1768639
 Number of obs: 73421; levels of grouping factors: 1128, 2972, 14

  Fixed-effects parameters:
─────────────────────────────────────────────────────
                  Coef.  Std. Error       z  Pr(>|z|)
─────────────────────────────────────────────────────
(Intercept)   3.27765     0.0235026  139.46    <1e-99
service      -0.0507441   0.0439855   -1.15    0.2486
─────────────────────────────────────────────────────
\end{verbatim}

\subsection{MovieLens Runtimes}\label{sec-app-ml}

The run times for the MovieLens models are summarized in
Table~\ref{tbl-sizespeed}. The movie cutoff is the threshold on the
number of ratings per movie; the user cutoff is the threshold on the
number of ratings per user. The various ns are the number of ratings,
users and movies in the resulting trimmed data set; the remaining
columns summarize the memory and time requirements of the associated
model.

\ifXeTeX
\footnotesize
\else
\small
\fi

\begin{longtable}[]{@{}
  >{\raggedleft\arraybackslash}p{(\linewidth - 18\tabcolsep) * \real{0.1161}}
  >{\raggedleft\arraybackslash}p{(\linewidth - 18\tabcolsep) * \real{0.1097}}
  >{\raggedleft\arraybackslash}p{(\linewidth - 18\tabcolsep) * \real{0.0839}}
  >{\raggedleft\arraybackslash}p{(\linewidth - 18\tabcolsep) * \real{0.0710}}
  >{\raggedleft\arraybackslash}p{(\linewidth - 18\tabcolsep) * \real{0.0774}}
  >{\raggedleft\arraybackslash}p{(\linewidth - 18\tabcolsep) * \real{0.1097}}
  >{\raggedleft\arraybackslash}p{(\linewidth - 18\tabcolsep) * \real{0.1161}}
  >{\raggedleft\arraybackslash}p{(\linewidth - 18\tabcolsep) * \real{0.0774}}
  >{\raggedleft\arraybackslash}p{(\linewidth - 18\tabcolsep) * \real{0.0903}}
  >{\raggedleft\arraybackslash}p{(\linewidth - 18\tabcolsep) * \real{0.1484}}@{}}
\caption{Summary of results for the \texttt{ml-32m} data modeling run
speeds}\label{tbl-sizespeed}\tabularnewline
\toprule\noalign{}
\begin{minipage}[b]{\linewidth}\raggedleft
\textbf{movie cutoff}
\end{minipage} & \begin{minipage}[b]{\linewidth}\raggedleft
\textbf{user cutoff}
\end{minipage} & \begin{minipage}[b]{\linewidth}\raggedleft
\textbf{ratings}
\end{minipage} & \begin{minipage}[b]{\linewidth}\raggedleft
\textbf{users}
\end{minipage} & \begin{minipage}[b]{\linewidth}\raggedleft
\textbf{movies}
\end{minipage} & \begin{minipage}[b]{\linewidth}\raggedleft
\textbf{model (GiB)}
\end{minipage} & \begin{minipage}[b]{\linewidth}\raggedleft
\textbf{L{[}2,2{]} (GiB)}
\end{minipage} & \begin{minipage}[b]{\linewidth}\raggedleft
\textbf{n eval}
\end{minipage} & \begin{minipage}[b]{\linewidth}\raggedleft
\textbf{time (s)}
\end{minipage} & \begin{minipage}[b]{\linewidth}\raggedleft
\textbf{time per eval (s)}
\end{minipage} \\
\midrule\noalign{}
\endfirsthead
\toprule\noalign{}
\begin{minipage}[b]{\linewidth}\raggedleft
\textbf{movie cutoff}
\end{minipage} & \begin{minipage}[b]{\linewidth}\raggedleft
\textbf{user cutoff}
\end{minipage} & \begin{minipage}[b]{\linewidth}\raggedleft
\textbf{ratings}
\end{minipage} & \begin{minipage}[b]{\linewidth}\raggedleft
\textbf{users}
\end{minipage} & \begin{minipage}[b]{\linewidth}\raggedleft
\textbf{movies}
\end{minipage} & \begin{minipage}[b]{\linewidth}\raggedleft
\textbf{model (GiB)}
\end{minipage} & \begin{minipage}[b]{\linewidth}\raggedleft
\textbf{L{[}2,2{]} (GiB)}
\end{minipage} & \begin{minipage}[b]{\linewidth}\raggedleft
\textbf{n eval}
\end{minipage} & \begin{minipage}[b]{\linewidth}\raggedleft
\textbf{time (s)}
\end{minipage} & \begin{minipage}[b]{\linewidth}\raggedleft
\textbf{time per eval (s)}
\end{minipage} \\
\midrule\noalign{}
\endhead
\bottomrule\noalign{}
\endlastfoot
1 & 20 & 32000204 & 200948 & 84432 & 56.25 & 53.11 & 26 & 18740.1 &
667.46 \\
2 & 20 & 31981597 & 200948 & 65825 & 35.41 & 32.28 & 19 & 7091.55 &
355.21 \\
5 & 20 & 31921467 & 200948 & 43884 & 17.47 & 14.35 & 26 & 3816.29 &
142.37 \\
10 & 20 & 31842705 & 200948 & 31961 & 10.72 & 7.61 & 25 & 2165.85 &
82.71 \\
15 & 20 & 31777786 & 200948 & 26428 & 8.31 & 5.2 & 24 & 1619.62 &
64.59 \\
20 & 20 & 31725920 & 200948 & 23350 & 7.16 & 4.06 & 21 & 1253.51 &
57.18 \\
50 & 20 & 31498689 & 200947 & 16034 & 4.99 & 1.92 & 28 & 1212.59 &
42.09 \\
1 & 40 & 30433400 & 144848 & 84205 & 55.8 & 52.83 & 23 & 15952.8 &
663.37 \\
2 & 40 & 30415014 & 144848 & 65819 & 35.25 & 32.28 & 32 & 11565.4 &
356.29 \\
5 & 40 & 30355623 & 144848 & 43884 & 17.31 & 14.35 & 38 & 5496.85 &
139.71 \\
10 & 40 & 30277758 & 144848 & 31961 & 10.56 & 7.61 & 21 & 1840.27 &
83.44 \\
15 & 40 & 30213583 & 144848 & 26428 & 8.15 & 5.2 & 21 & 1428.19 &
64.85 \\
20 & 40 & 30162330 & 144848 & 23350 & 7.0 & 4.06 & 17 & 1026.09 &
57.08 \\
50 & 40 & 29938038 & 144848 & 16034 & 4.83 & 1.92 & 24 & 1046.22 &
41.57 \\
1 & 80 & 27569316 & 94380 & 83897 & 55.13 & 52.44 & 23 & 16155.2 &
677.07 \\
2 & 80 & 27551218 & 94380 & 65799 & 34.94 & 32.26 & 22 & 8541.43 &
392.76 \\
5 & 80 & 27492886 & 94380 & 43884 & 17.03 & 14.35 & 23 & 3564.96 &
142.58 \\
10 & 80 & 27416310 & 94380 & 31961 & 10.28 & 7.61 & 23 & 2054.17 &
85.29 \\
15 & 80 & 27353366 & 94380 & 26428 & 7.87 & 5.2 & 26 & 1785.08 &
65.72 \\
20 & 80 & 27303019 & 94380 & 23350 & 6.72 & 4.06 & 43 & 2527.21 &
57.72 \\
50 & 80 & 27083536 & 94380 & 16034 & 4.55 & 1.92 & 20 & 873.98 &
41.38 \\
\end{longtable}

\normalsize
\clearpage

\subsection{Dimensions of quantities in linear mixed
models}\label{sec-re-dim}

The following table is mostly a reproduction of Table 3 from
\citet{bates.maechler.etal:2015}, with slight modifications.

\begin{longtable}[]{@{}
  >{\raggedright\arraybackslash}p{(\linewidth - 4\tabcolsep) * \real{0.1800}}
  >{\raggedright\arraybackslash}p{(\linewidth - 4\tabcolsep) * \real{0.6200}}
  >{\raggedright\arraybackslash}p{(\linewidth - 4\tabcolsep) * \real{0.2000}}@{}}
\caption{Sizes of model components}\label{tbl-re-sizes}\tabularnewline
\toprule\noalign{}
\begin{minipage}[b]{\linewidth}\raggedright
Symbol
\end{minipage} & \begin{minipage}[b]{\linewidth}\raggedright
Description
\end{minipage} & \begin{minipage}[b]{\linewidth}\raggedright
\texttt{insteval} example
\end{minipage} \\
\midrule\noalign{}
\endfirsthead
\toprule\noalign{}
\begin{minipage}[b]{\linewidth}\raggedright
Symbol
\end{minipage} & \begin{minipage}[b]{\linewidth}\raggedright
Description
\end{minipage} & \begin{minipage}[b]{\linewidth}\raggedright
\texttt{insteval} example
\end{minipage} \\
\midrule\noalign{}
\endhead
\bottomrule\noalign{}
\endlastfoot
\(n\) & length of the response vector \(\mcY\) & \(n = 73{,}421\) \\
\(p\) & no. of fixed effects in the model matrix \(\mbfX\) &
\(p = 2\) \\
\(k\) & no. of random effects terms & \(k = 3\) (\texttt{s}, \texttt{d},
\texttt{dept}) \\
\(p_i\) & no. of fixed effects terms in the \(i^\text{th}\) random
effect & \((1,1,2)\) \\
\(\ell_i\) & no. of levels of the \(i^\text{th}\) grouping factor &
\((2972, 1128, 14)\) \\
\(q_i = p_i \ell_i\) & no. of columns of the term-wise model matrix &
\((2972, 1128, 28)\) \\
\(q = \sum_{i}q_i\) & no. of columns in the random effects model matrix
\(\mbfZ\) & \(q = 4{,}128\) \\
\(q + p + 1\) & no. of columns (and rows) of the augmented matrix
\(\mbfA\) & \(4131\) \\
\end{longtable}

\subsection{Computing environment}\label{sec-app-env}

The computing environment for all the timings, except for the models fit
to the \texttt{ml-32m} data, was

\begin{Shaded}
\begin{Highlighting}[]
\FunctionTok{versioninfo}\NormalTok{()}
\end{Highlighting}
\end{Shaded}

\begin{verbatim}
Julia Version 1.11.5
Commit 760b2e5b739 (2025-04-14 06:53 UTC)
Build Info:
  Official https://julialang.org/ release
Platform Info:
  OS: macOS (arm64-apple-darwin24.0.0)
  CPU: 11 × Apple M3 Pro
  WORD_SIZE: 64
  LLVM: libLLVM-16.0.6 (ORCJIT, apple-m2)
Threads: 5 default, 0 interactive, 2 GC (on 5 virtual cores)
Environment:
  JULIA_PROJECT = @.
  JULIA_LOAD_PATH = @:@stdlib
\end{verbatim}

with an accelerated BLAS library

\begin{Shaded}
\begin{Highlighting}[]
\NormalTok{BLAS.}\FunctionTok{get\_config}\NormalTok{()}
\end{Highlighting}
\end{Shaded}

\begin{verbatim}
LinearAlgebra.BLAS.LBTConfig
Libraries: 
├ [ LP64] Accelerate└ [ILP64] Accelerate
\end{verbatim}

The versions of the Julia packages used are

\begin{Shaded}
\begin{Highlighting}[]
\ImportTok{using} \BuiltInTok{Pkg;}\NormalTok{ Pkg.status()}
\end{Highlighting}
\end{Shaded}

\begin{verbatim}
Status `~/projects/BlockedCholeskyMM/Project.toml`
  [cbdf2221] AlgebraOfGraphics v0.10.5
  [13e28ba4] AppleAccelerate v0.4.0
  [69666777] Arrow v2.8.0
  [336ed68f] CSV v0.10.15
  [13f3f980] CairoMakie v0.13.5
  [0ca39b1e] Chairmarks v1.3.1
  [a93c6f00] DataFrames v1.7.0
  [33e6dc65] MKL v0.8.0
  [ff71e718] MixedModels v4.35.0
  [08abe8d2] PrettyTables v2.4.0
  [856f044c] MKL_jll v2025.0.1+1
  [de0858da] Printf v1.11.0
  [2f01184e] SparseArrays v1.11.0
\end{verbatim}

\end{document}